%% file: main.tex
\documentclass[journal]{IEEEcsmag}
\usepackage[colorlinks,urlcolor=blue,linkcolor=blue,citecolor=blue]{hyperref}
\expandafter\def\expandafter\UrlBreaks\expandafter{\UrlBreaks\do\/\do\*\do\-\do\~\do\'\do\"\do\-}
\usepackage{upmath,color}
\usepackage[utf8]{inputenc}
\usepackage{amsmath,amsthm,mathtools,tikz}
\usepackage{amsrefs,amssymb}
\usepackage{algorithm2e}
\usepackage{ulem}
\usepackage{float}
\usepackage{comment}
\usepackage{subfiles}
\usepackage{subfigure}
\usepackage{svg} 
\graphicspath{{\subfix{figures/}}} 

\sptitle{Feature Article}

 \title{In-situ data extraction for pathway analysis in an idealized atmosphere configuration of E3SM}

\author{Andrew Steyer}
\affil{Sandia National Laboratories, Albuquerque, NM, USA}

\author{\text{Luca} Bertagna}
\affil{Sandia National Laboratories, Albuquerque, NM, USA}

\author{Graham Harper}
\affil{Sandia National Laboratories, Albuquerque, NM, USA}

\author{Jerry Watkins}
\affil{Sandia National Laboratories, Livermore, CA, USA}

\author{Irina Tezaur}
\affil{Sandia National Laboratories, Livermore, CA, USA}

\author{Diana Bull}
\affil{Sandia National Laboratories, Albuquerque, NM, USA  }

\newcommand{\clderatools}{\texttt{CLDERA-Tools}}

\newcommand{\esm}{\texttt{E3SM}}
\newcommand{\esmvii}{\texttt{E3SMv2}}
\newcommand{\esmvi}{\texttt{E3SMv1}}

\newcommand{\ct}{\texttt{CLDERA-Tools}}

\newcommand{\ikt}[2]{{\color{red}{#1}} {\color{red}{\sout{#2}}}}

\begin{document}

\markboth{FEATURE}{FEATURE}

\begin{abstract}
We propose an approach for characterizing source-impact \textit{pathways}, the interactions of a set of variables in space-time due to an external forcing, in climate models using in-situ analyses that circumvent computationally expensive read/write operations.  This approach makes use of a lightweight open-source software library we developed known as \clderatools{}.  We describe how \clderatools{} is linked with the U.S. Department of Energy's Energy Exascale Earth System Model (\esm{}) in a minimally invasive way for in-situ extraction of quantities of interested and associated statistics.  Subsequently, these quantities are used to represent source-impact pathways with time-dependent directed acyclic graphs (DAGs). The utility of \clderatools{} is demonstrated by using the data it extracts in-situ to compute a spatially resolved DAG from an idealized configuration of the atmosphere with a parameterized representation of a volcanic eruption known as HSW-V.    
\end{abstract}

\maketitle

\section{Introduction}
\label{section_introduction}

\subfile{sections/1_introduction.tex}

\section{Climate modeling preliminaries}
\label{climate_modeling}
\subfile{sections/2_climatemodeling.tex}

\section{Physical pathways in climate simulations}
\label{section_pathways}

\subfile{sections/3_pathways.tex}

\section{In-situ extraction of climate data}\label{section_insitu}

\subfile{sections/4_insitu.tex}
\section{Results of Numerical Experiments}
\label{section_results}

\subfile{sections/5_results.tex}

\section{Conclusion}
\label{section_conclusion}

\subfile{sections/6_conclusion.tex}

\section*{Acknowledgements}

The authors would like to thank Joe Hollowed and Christiane Jablonowski for valuable discussions and feedback related to the HSW-V model.  This work was supported by the Sandia National Laboratories Laboratory Directed Research \& Development Grand Challenge Program.  Sandia National Laboratories is a multimission laboratory managed and operated by National Technology and Engineering Solutions of Sandia, LLC., a wholly owned subsidiary of Honeywell International, Inc., for the U.S. Department of Energy's National Nuclear Security Administration under grant~DE-NA-0003525.  This paper describes objective technical results and analysis.  Any subjective views or opinions that might be expressed in the paper do not necessarily represent the views of the U.S. Department of Energy or the United States Government.

\bibliography{cldera_bib}

\end{document}

%% file: sections/1_introduction.tex
As impacts in the climate system become more acute, there is increasing interest in attributing the cause of the impacts.  Most attribution studies in the climate are global in nature and tend to focus on anthropogenic climate change \cite{RN1}.  However, there is a need to understand how localized disturbances in the climate (e.g., volcanoes, large wildfires, atmospheric rivers, tipped systems) may themselves result in climate effects. We expand this understanding by developing algorithms to represent source-impact pathways, the relationships and interactions of a set of climate variables in space-time due to an external forcing, from data that is extracted in-situ from a simulated volcanic eruption. 


In this paper, we develop a mathematical framework and associated algorithm (Algorithm \ref{alg:pathwayalg}) for representing pathways from high-frequency time-series of model quantities of interest (QOIs) as time-dependent directed acyclic graphs (DAGs).  In this framework, relationships between variables are defined by scientific subject matter expertise and physics and an impact of a variable due to a forcing is measure by comparing its forced value to its baseline unforced statistics with a bounds test.  Thereafter, we describe a software package, \ct{}, for high-frequency in-situ tracking of statistics of multiple QOIs in the  U.S. Department of Energy's Energy Exascale Earth System Model (\esm{}) \cite{e3smv2}.  We demonstrate our mathematical framework for pathways by using \ct{} to extract statistics of several climate variables from an idealized climate configuration in \esm{} that is subject to forcing from a volcanic eruption modeled as an injection of SO2 aerosol \cite{hollowedetal2024}.  In particular, we show the efficiency of the in-situ QOI tracking in \ct{} and how the pathways computed using Algorithm \ref{alg:pathwayalg} change with respect to the eruption mass size and sensitivity of the defined bounds tests.

The rest of the paper is structured as follows. In Section \ref{climate_modeling}, we describe the \esm{} climate model and the simplified atmosphere model configuration that is used to test our pathway detection algorithm. In Section \ref{section_pathways}, we give a concrete definition of pathways, develop an algorithm to compute pathways for a fully-discrete forward model, and describe how we implement this algorithm for the simplified atmosphere configuration we considered. In Section \ref{section_insitu} we describe the \ct{} software package for in-situ analysis, including how it is linked to \esm{} and how it computes statistics of QOIs. Section 5 presents the results of using \ct{} on a stratospheric aerosol injection 
temperature pathway problem, including a cost study to show the efficiency of our in-situ approach. Finally, in Section \ref{section_conclusion}, we draw some conclusions and discuss perspectives on future work.

%% file: sections/2_climatemodeling.tex
\subsection{The atmosphere component of \esmvii{}}\label{sec:eam}


\esmvii{} \cite{e3smv2} is the second version of \esm{} that offers improved computational efficiency and climate fidelity over the first version \esmvi{}. It is comprised of several component models for the atmosphere, ocean, land, and cryosphere that can be coupled in various resolutions and configurations.

The E3SM Atmosphere Model (EAM) is the atmosphere component model of \esmvii{}. Broadly speaking, EAM is divided in two parts: dynamics (``dycore") and physics parametrizations (``physics"). Essentially, the dycore solves the Navier-Stokes equations and transports passive tracers, while physics approximate the effect of physics processes (e.g., radiation, aerosols, clouds) that are unresolved by dynamics.

The computational grid for dynamics consists of a quadrilateral horizontal mesh, vertically extruded for a certain number of vertical layers. EAM uses spectral elements methods in the horizontal direction and finite difference schemes in the vertical direction \cite{hannahetal2021}. Physics are resolved on a computational grid obtained by subdividing each horizontal quadrilateral into four finite volume cells, again extruded vertically.  Such grids are referred to as ``cubed sphere'' grids, since the horizontal grid is obtained from a uniformly meshed cubed surface, inflated into a sphere, and are denoted as neX, where X is the number of elements along each cube edge (see \cite{hannahetal2021}). 
  



\subsection{The modified HSW-V configuration}\label{sec:modhsw}

To reduce the computational cost associated with running \esmvii{} and isolate the climatological effects due to volcanic eruptions from internal variability, we adopt a modified Held-Suarez-Williamson (HSW) atmosphere-only configuration \cite{williamson1998}. A detailed description of this configuration, referred to as HSW-V, is contained in \cite{hollowedetal2024}. 

The HSW-V configuration adds idealized analytic physics and chemistry parameterizations that simulate the chemistry, radiative, and temperature effects of sulfur dioxide and ash resulting from a volcanic eruption.  Specifically, it adds the capability to: (i) inject SO2 and ash tracers into the stratosphere, (ii) have the SO2 react via a simple chemistry relation into sulfate (SO4), and (iii) differentially adjust the radiative forcing in a given region based on the total tracer content (represented by aerosol optical depth) such that the stratosphere warms and the surface cools.  The atmosphere temperature is constantly nudged towards a daytime equilibrium temperature in a dry, aseasonal, and topography free environment.  Although the equilibrium temperature profile in HSW-V is perturbed analytically by the aerosol optical depth field, the temperature field itself still has internal variability due to sensitive dependence on initial conditions. Therefore, signal-to-noise issues are still present within the dataset for the temperature QOIs which motivates our ensemble analysis of pathways in Section \ref{sec:pathwayanalysis}.

The results in Section \ref{section_results} use HSW-V on an ne16 cubed-sphere grid (about $2^{\circ}$ horizontal resolution), with 72 vertical layers in an approximately 60km thick atmosphere.  Although this grid is relatively coarse, its resolution is sufficient to recover the correct stratospheric warming and tropospheric cooling associated with our target application of computing the temperature pathway of the Mount Pinatubo eruption. We utilize the limited-variability ensemble approach in \cite{hollowedetal2024} with an eruption occurring day 90 of the simulation.




%% file: sections/3_pathways.tex
Heuristically, a source-impact pathway in a forward model is the representation of the relationships and interactions between a set of QOIs in space and time when the model is subjected to an external forcing.  Mathematically, we represent pathways using DAGs: nodes represent QOIs and edges represent relationships between QOIs (the direction of the edges denotes the flow of impacts and is determined by subject matter expertise and physics).




\subsection{Algorithm to construct pathway DAGs}
  
Consider a forward model of the following form:
\begin{equation}\label{eq:forwardmodel}
u_{m+1} = F(u_{m},t_m,\alpha), \quad u_{m} \in \mathbb{R}^J \times \mathbb{R}^d, \quad \alpha \in \mathbb{R}^p,
\end{equation}
where $m=1,\hdots,M$, $t_m$ represents the model time at time-step $m$, $u_{m}$ represents the model state at time-step $m$, and $\alpha$ represents the model parameters. We will also denote with $u_{m,j,d}$ the state $u_m$ of quantity $d = 1,\hdots,D$ at the spatial location $j=1,\hdots,J$. Associated to \eqref{eq:forwardmodel} is a set of $r$ QOIs $Q_{1},\hdots,Q_{r}$ where each $Q_{l}:\mathbb{R}^J \times \mathbb{R}^d \rightarrow \mathbb{R}$ is a function of the model state $u_m$ where we use the notation $Q_{l,m} := Q_l(u_m)$. 

Our approach for constructing a pathway DAG based on \eqref{eq:forwardmodel} begins with what we call the \text{base-DAG}.  The base-DAG $\mathcal{G}_B=(V_B,E_B)$ consists of a set of $r$-vertices $V=\{v_l\}_{l=1}^{r}$ with node $v_{l}$ corresponding to the QOI $Q_l$ and a set of edges $E = \{e_j\}_{j=1}^{s}$ determined by assumed relationships between QOIs.  These hypothesized relationships are informed by subject matter expertise (e.g., increases in tracer concentrations causes increases in aerosol optical depth).  Associated to each node $v_{l}$ is a (potentially time-dependent) bounds test $\tau_{l,m}$, taking Boolean values in $\{0,1\}$, that determines when a node will be active ($\tau_{l,m} = 1$) or inactive ($\tau_{l,m} = 0$). See Equations \ref{eq:so2boundstest}-\ref{eq:aodboundstests} in Section \ref{sec:pathwayanalysis} for concrete examples of the bounds tests we use.

We shall compute a \text{pathway-DAG} $\mathcal{G}$ defined by a finite sequence $\mathcal{G} = \{\mathcal{G}_m\}_{m=0}^{M}$ such that each $\mathcal{G}_m = (V_m,E_m)$ is a DAG with set of vertices $V_m \subseteq V$ and set of edges $E_m \subseteq E$.  The following algorithm is used to compute the pathway-DAG using the convention that $E_{-1} = \emptyset$.
\begin{algorithm}\label{alg:pathwayalg}
\caption{Algorithm to compute pathway-DAG}\label{alg:two}
\SetKwInOut{Input}{Input}
\SetKwInOut{Output}{Output}
\Input{$\mathcal{G}_B=(V_B,E_B)$, $\{u_m\}_{m=0}^{M}$, $\{Q_l\}_{l=1}^{r}$, $\{\tau_{l,m}\}_{l=1,m=0}^{r,M}$}
\Output{$\mathcal{G}=\{\mathcal{G}_m\}_{m=0}^{M}$}
\For{$m=0,\hdots,M$}{

  Compute $Q_{1,m},\hdots,Q_{r,m}$
  
  Compute $\tau_{1,m},\hdots,\tau_{r,m}$

  Set $V_m = \cup_{l=1}^{r} \{v_l \in V_B: \tau_{l,m} = 1\}$; 

  Set $E_m = \cup_{j=1}^{s} \{e_j \in E_B: e_{j,1},e_{j,2} \in V_m \}$

  Set $\mathcal{G}_m = (V_m,E_m)$
}
\end{algorithm}

For each $m \geq 0$, Algorithm \ref{alg:pathwayalg} constructs the vertex set $V_m$ as the set of all vertices in the base-DAG vertex set $V_B$ with a bounds test value equal to $1$.  The edge set $E_m$ is constructed from $V_m$ by including all edges in the base-DAG edge set $V_B$ whose component vertices are in $V_m$. 

\subsection{Pathway computation in HSW-V}\label{sec:pathwayhsw}

\begin{figure}
   \centering
   \includegraphics[width=1.8\linewidth]{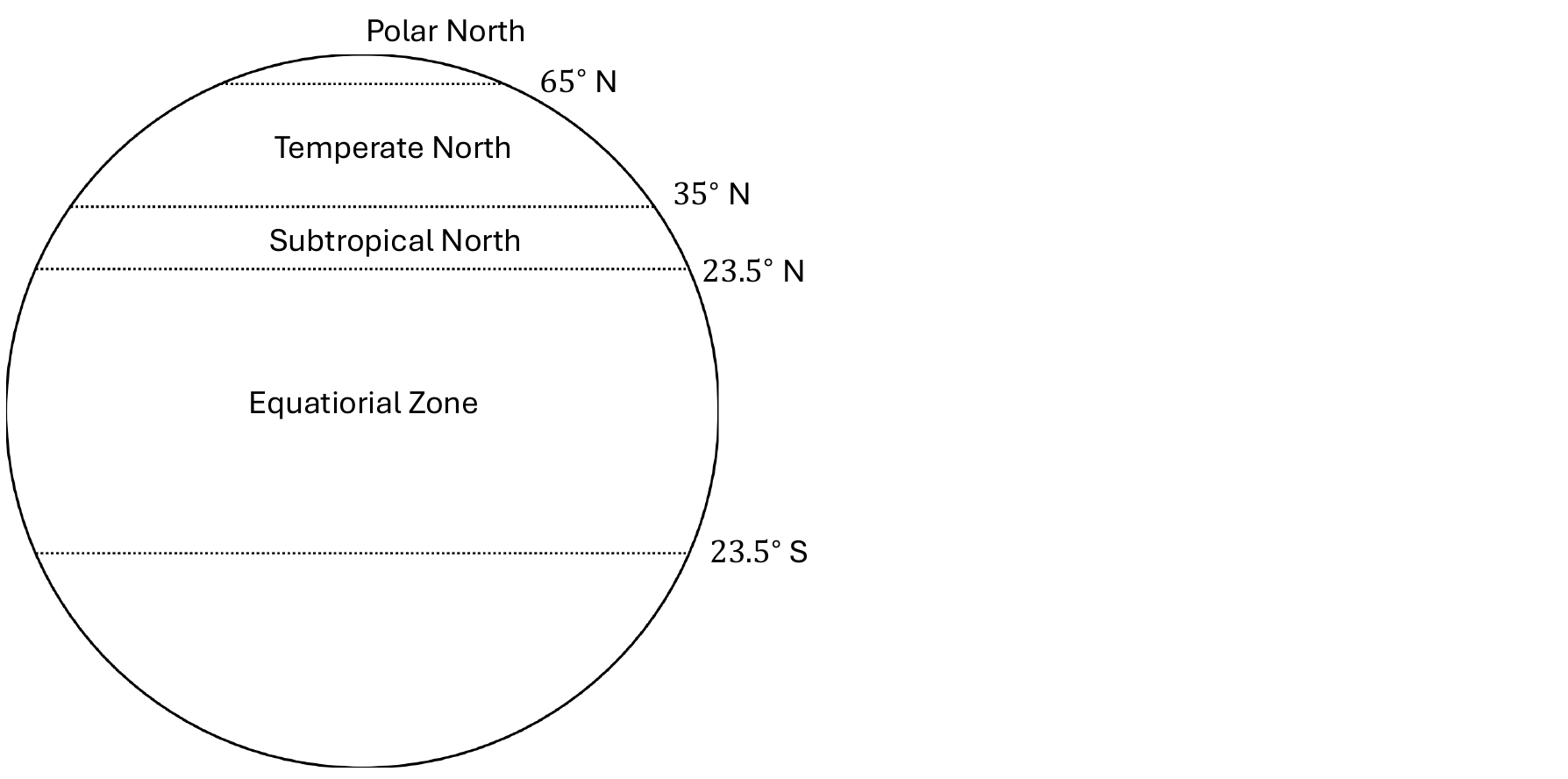}\\
   \centering
 \includegraphics[width=0.9\linewidth]{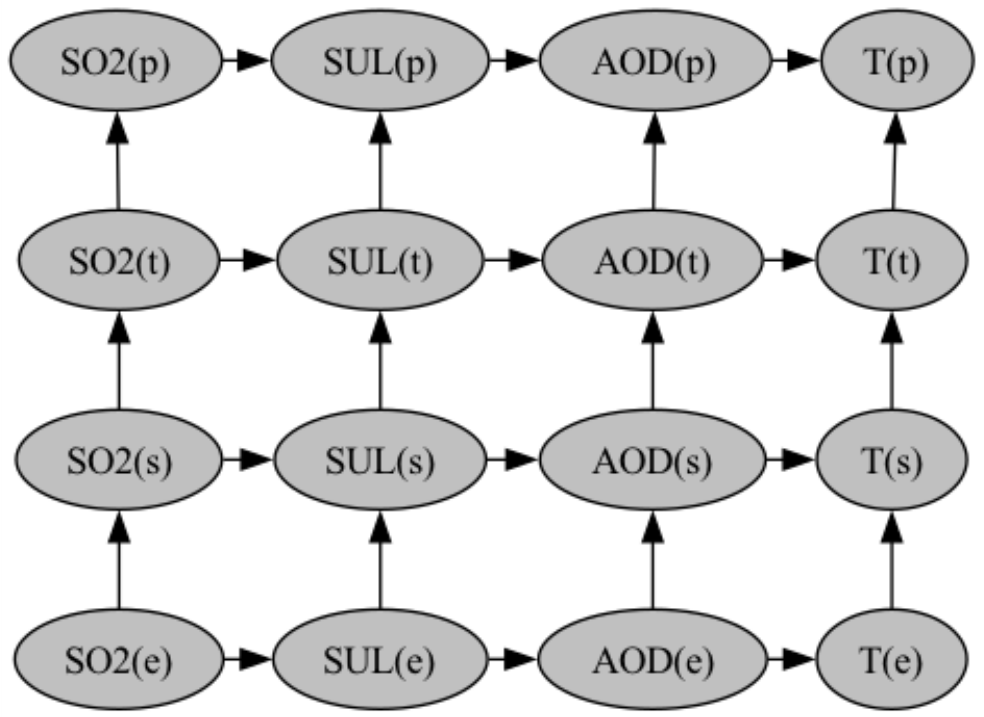}
    \caption{(Top) Zonal regions used to define QOIs in the HSW-V configuration.  (Bottom) Base-dag for QOIs we define for the HSW-V configuration where the names of the nodes are defined in the first paragraph of Section \ref{sec:pathwayhsw}. }
    \label{fig:basedagetc}
\end{figure}

We consider eruption effects in HSW-V, focusing on the northern hemisphere and the equatorial zone.  We subdivide this area into four latitudinal bands:  the equatorial zone (bounded between the latitudes $23.5^{\circ}$ S and $23.5^{\circ}$ N and denoted by ``e"),
the subtropical north zone (bounded between the latitudes $23.5^{\circ}$ N and $35^{\circ}$ N and denoted by ``s"),
the temperate north zone (bounded between the latitudes $35^{\circ}$ N  and $66.5^{\circ}$ N and denoted by ``t"),
and the polar north zone (bounded between the latitudes $66.5^{\circ}$ N and $90^{\circ}$ N and denoted by ``p").  Figure \ref{fig:basedagetc} (top) offers a visualization of these zonal regions.  We consider four model fields:  stratospheric SO2 and sulfate (SUL) concentration, aerosol optical depth (AOD), and stratospheric temperature (T). At each model time-step, we extract the SO2, SUL, and temperature fields from the model, and
then compute a pressure-weighted vertical integral over model levels that correspond to the mid-stratosphere using pressures in the approximate
range of 25-75 hPa, and subsequently compute an area-weighted integral of the result over each zone (equatorial, temperate, subtropical, and polar).
For the AOD field, we only compute an area-weighted integral over each zone since it is a two-dimensional field.
The QOIs we consider for the remainder of this paper are these integrated model fields, where for each model field Q (Q=SO2,SUL,AOD,T) we denote by Q(x) (x=e,s,t,p) the integrated value of Q in zone x. 

The base-DAG for these QOIs is shown in Figure \ref{fig:basedagetc} (bottom) and was determined as follows.  The subgraphs $\text{SO2(x)}\rightarrow \text{SUL(x)} \rightarrow \text{AOD(x)}\rightarrow \text{T(x)}$, where $\text{x}=\text{e,s,t,p}$ is a given latitudinal band, correspond to the physical process defined in the HSW-V configuration where SO2 reacts to form SO4 (which reduces gradually over time) which in turn increases the AOD, which then impacts temperature by affecting atmospheric heating.  The subgraphs $\text{Q(e)}\rightarrow \text{Q(s)} \rightarrow \text{Q(t)}\rightarrow \text{Q(p)}$ where $\text{Q}=\text{SO2,SUL,AOD,T}$ correspond to the Brewer-Dobson circulation in the stratosphere in which tropical air masses are transported poleward.  


%% file: sections/4_insitu.tex

We identify two main approaches for implementing in-situ coanalysis for \esm{} (or any other forward model):
\begin{itemize}
  \item Directly modify the existing model's codebase to handle computing, storing, and processing of QOIs.
  \item Link the model to a secondary software package that handles computing, storing, and processing of QOIs.
\end{itemize}
The first approach (directly modifying the model) has the advantage of providing direct access to the model's internal structures, but comes at a potential cost of extensive changes to an already complex codebase. Costs and benefits are swapped for the second approach (linking to a secondary software package): changes to the model codebase will be limited to a handful of function callbacks, but additional care has to be paid to build and link the secondary library correctly.

We have chosen the second approach for the following reasons. First, minimizing modifications in \esm{} makes it much easier to update the model as new versions are released. Second, a separate and self-contained package can potentially be used for in-situ coanalysis with different models.
Third, a small self-contained package makes unit-testing easier since it avoids the rigidity of a complex build system.  Finally, adding dependencies on third-party libraries is much easier to handle in a small library than in a complex code base such as \esm{}.

We have implemented an in-situ coanalysis open-source software package called \ct{}. Although \ct{} was developed with \esm{} as a target application, it has enough flexibility to be easily adapted for in-situ coanalysis in another forward model. The code base of \ct{} is divisible into a few macro areas:

\begin{itemize}
  \item Model interfaces: C as well as Fortran interfaces that can be called from the forward model to setup and run coanalysis.
  \item QOIs: a set of general-purpose QOIs, such as global/regional averages, vertical integrals, masked integrals, etc., all of which can be ``piped", so that the output of a QOI can be the input of another.
  \item Input-output (I/O): a simple interface to pnetcdf, that can be used to efficiently save the results of the coanalysis to disk.
  \item In-memory storage: a collection of data structures used to store input and output data as the simulation progresses.
\end{itemize}
Among the in-memory data storage structures, the \texttt{Field} class is of particular importance, since it allows sharing input data with the
forward model. A field is a collection of metadata (name, layout, etc.) along with a pointer to the actual data.  To avoid data movement, the field can be constructed from raw pointers from the model (if available) during initialization so that no additional data transfer is needed at run time. In EAM, most model variables are stored in persistent memory, so that we could construct field objects in \ct{} that simply ``view'' the existing model data. However, certain quantities are computed on-the-fly in EAM, without being stored in persistent memory. For this reason, \ct{} can also create fields as self-contained allocations where the data has to be copied over during run time.

The typical use of \ct{} is divisible into an initialization and a runtime phase. During initialization, the following steps happen:

\begin{itemize}
  \item \ct{} reads in a list of all the QOIs to compute and analyze, along with the desired I/O configuration.
  \item the forward model passes to \ct{} a pointer for all its variables; \ct{} will wrap them in \texttt{Field} objects, for later usage;
  \item \ct{} creates and sets up all the structures needed for QOIs calculation;
  \item I/O is initialized and output files are readied for writing operations.
\end{itemize}

The initialization step is where most of the callback functions to \ct{} must be inserted in the forward model. Once that is done, at runtime, the model only needs to call one interface from \ct{} to compute all the QOIs that were requested at initialization. In this phase, \ct{} will compute each QOI, perform time-averaging (if requested), and call the I/O layer to save the computed QOIs.

Our efficient implementation of the calculation of QOIs in \ct{} allows our in-situ approach to use a higher time resolution than it would be possible with native model output. Thus, in-situ coanalysis can be viewed as a tool for accessing and processing high-resolution and high-volume data, without requiring expensive read/write operations or excessive disk storage.

The process of connecting \esm{} to \ct{} is relatively straightforward, but illustrating it requires detailed knowledge of the \esm{} build system, which is beyond the scope of this paper. Nevertheless, the reader can verify the minimal invasiveness of the code modifications, by accessing our public fork of \esm{} \cite{cldera-e3sm}, and searching for preprocessor strings of the form \verb|#ifdef CLDERA_PROFILING| in the EAM component.

%% file: sections/5_results.tex
In this section, we present results from several computational experiments showing that \ct{} is efficient for in situ coanalysis and how we use this data for pathway analysis using Algorithm \ref{alg:pathwayalg}.  


\subsection{Cost Study}\label{sec:coststudy}


In-situ coanalysis software such as \ct{} must be efficient to be useful.  To show that this is the case for pathway analysis, we present some timings for running \esm{} linked with \ct{} for in-situ tracking with different numbers of QOIs. These experiments were run on a single node of a local cluster, featuring a dual-socket Intel Xeon Gold processor with 18 cores per socket, 2 hardware threads per core, and 96GB of RAM per socket. The baseline results were obtained by taking the mean value of the CPL$:$ATM\_RUN E3SM \esm{} timer from a 10 member ensemble running \esm{} without \ct{} enabled.  The cost experiment results were obtained by taking the mean value of the same timer from a 10 member ensemble running \esm{} with \ct{} enabled tracking 7, 35, 175, or 875 QOIs. For all these runs, all QOIs were computing the zonal average of a vertically integrated variable, which is the typical (and most expensive) QOI that we used in our numerical experiments. 

\begin{figure}
    \centering
    \resizebox{8.0cm}{!}{\includegraphics{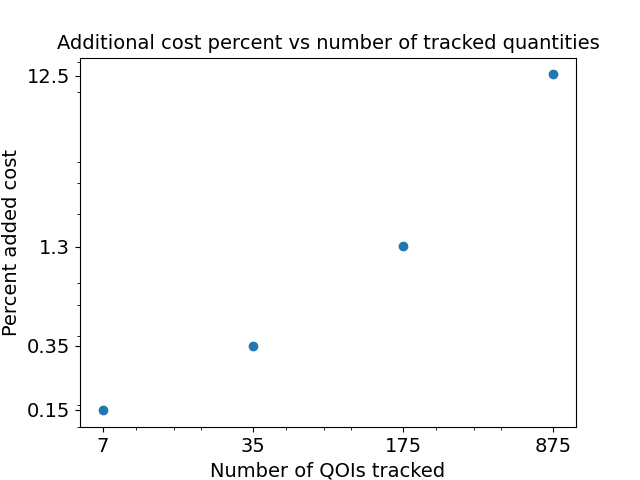}}
    \caption{The additional computational overhead of running in-situ tracking with \ct{} scales with the number of QOIs being tracked.}
    \label{fig: performance averages}
\end{figure}

The additional cost percent of \ct{} was determined by taking the ratio of the run-time of each experiment with the baseline. The results are shown in Figure \ref{fig: performance averages}.  For experiments running \esm{} with \ct{} when fewer than 175 QOIs are tracked, the additional cost for in-situ tracking is minimal ($<1.3\%$).  However, in-situ tracking of a larger number of QOIs ($>175$) can result in a nontrivial additional cost of more than $10\%$, which may be considered acceptable depending on the computational resources available and application.  The pathway results in the next subsection required tracking $16$ QOIs resulting in minimal additional computational overhead.
 
\subsection{Pathway Analysis}\label{sec:pathwayanalysis}

In this section we present the results of several experiments using Algorithm \ref{alg:pathwayalg} to compute pathways from \esm{} data extracted in-situ using \ct{}with the QOIs and base-DAG specified in Section \ref{sec:pathwayhsw}.  Use of Algorithm \ref{alg:pathwayalg} requires specifying bounds tests ($\tau_{l,m}$ in Algorithm \ref{alg:pathwayalg}) for each QOI.  We let SO2(x)$_m$, SUL(x)$_m$, AOD(x)$_m$, and T(x)$_m$ denote the value of each QOI at time-step $m$. To define tests for the temperature QOIs T(x) ($x$=e,s,t,p), we use an ensemble of eruption-free HSW-v runs (no ash or SO2 is injected) to define baseline statistics T$_{\mu}$(x), the mean value of T(x) in the eruption-free ensemble, and T$_\sigma$(x), the value of the standard deviation of the eruption free ensemble. We let T$_{\mu}$(x)$_m$ and T$_\sigma$(x)$_m$ denote, respectively, the values of T$_{\mu}$(x) and T$_\sigma$(x) at time-step $m$.  In a run where a nonzero amount of SO2 and ash is injected, we compute of T$_Z$(x)$_m$=(T(x)$_m$-T$_{\mu}$(x)$_m$)/T$_\sigma$(x)$_m$.  We specify upper and lower bounds ($T_{u} >0$ and $T_{l} \leq T_u$, respectively) and then define a bounds test $\tau_{\text{T(x)},m}$ for T(x)$_m$ as follows:
\begin{equation}\label{eq:Tboundstest}
\tau^{\text{T(x)}}_{m} = \left\{\begin{array}{cc} 
0 & m = 0 \\
0 &  \text{T}_Z\text{(x)}_m \leq T_{l}, m > 0\\
1 &  \text{T}_Z\text{(x)}_m \geq T_{u}, m > 0 \\
\tau^{\text{T(x)}}_{m-1} &   T_l \leq  \text{T}_Z\text{(x)}_m \leq T_u, m > 0.
\end{array} \right.
\end{equation}

Enforcing $T_u > T_l$ in in \eqref{eq:Tboundstest}  prevents the test from frequently alternating between 0 and 1 due to small fluctuations in $T_Z$.  To see this note that if $T_u=T_l=\overline{T}$ and $T_Z \approx \overline{T}$, then $T_Z$ may alternate between 0 and 1 as often as every model time-step. Setting $T_u > T_l$ sufficient large filters out noise from natural variability and better highlights the underlying signal. Note that the value $T_u - T_l$ characterizes the sensitivity of the bounds test, with smaller values denoting increased sensitivity of T(x) to being activated.

In an eruption-free HSW-V configuration, the values of SO2(x), SUL(x), and AOD(x) are zero for all times since no SO2 is present to be converted to sulfate and thereby increase the aerosol optical depth.  Therefore using the bounds test as for T(x) would not be well-defined, since the associated standard deviations of SO2(x), SUL(x), and AOD(x) for the eruption-free ensemble is identically zero.  With this in mind, we respectively define bounds tests $\tau^{\text{SO2(x)}}_{m}$, $\tau^{\text{SUL(x)}}_{m}$, $\tau^{\text{AOD(x)}}_{m}$ for  SO2(x), SUL(x), AOD(x) as follows:


\begin{equation}\label{eq:so2boundstest}
\tau^{\text{SO2(x)}}_m = \left\{\begin{array}{cc} 
0 &  \text{SO2(x)}_m \leq 4\times 10^{-10},\\
1 &  \text{SO2(x)}_m  \geq 8\times 10^{-10}, \\
\tau^{\text{SO2(x)}}_{m-1} &   \text{SO2(x)}_m  \in (4\times 10^{-10},8\times 10^{-10}).
\end{array} \right.
\end{equation}

\begin{small}
\begin{equation}\label{eq:sulboundstest}
\tau^{\text{SUL(x)}}_{m} = \left\{\begin{array}{cc} 
0 &  \text{SUL(x)}_m \leq 4\times 10^{-10},\\
1 &  \text{SUL(x)}_m  \geq 8\times 10^{-10}, \\
\tau^{\text{SUL(x)}}_{m-1} &   \text{SUL(x)}_m  \in(4\times 10^{-10}, 8\times 10^{-10}).
\end{array} \right.
\end{equation}
\small
\begin{equation}\label{eq:aodboundstests}
\tau^{\text{AOD(x)}}_{m} = \left\{\begin{array}{cc} 
0 &  \text{AOD(x)}_m \leq 0.0075,\\
1 &  \text{AOD(x)}_m  \geq 0.015, \\
\tau^{\text{AOD(x)}}_{m-1} &   \text{AOD(x)}_m \in(0.0075,0.015).
\end{array} \right.
\end{equation}
\end{small}

where the bounding values $4\times 10^{-10}$, $8\times 10^{-10}$ for SO2(x) and SUL(x) and $0.0075$, $0.015$ for AOD(x) were chosen based upon the size of the eruption and the expected concentrations of aerosol quantities.  The bounding values $T_u$ and $T_l$ are varied in several experiments to determine the sensitivity of Algorithm \ref{alg:pathwayalg} to this test.  

Each experiment is defined as follows.  We run an ensemble of 10 eruption runs with specified values of $T_u \geq T_l > 0$ and compute the associated pathway with Algorithm \ref{alg:pathwayalg} for each ensemble member.  Ensemble members have nearly identical initial conditions which differ only by a small random perturbation of the temperature field.  Note that the model is chaotic and this perturbation results in model trajectories that are  pointwise decorrelated over time, but with similar statistical  properties.  To test the sensitivity of the algorithm with respect to eruption size we run simulations with a 10 Tg mass eruption corresponding to the true Mount Pinatubo eruption size as well as a half (5 Tg) and double mass (20 Tg) eruption.  For each eruption mass ensemble we use $T_l = 0.5$ and we use labels Ex1,Ex2,Ex3,Ex4 for various experiments where Ex1 corresponds to $T_u = 0.75$, Ex2 to $T_u = 1.0$, Ex3 to $T_u = 1.5$, and Ex4 to $T_u = 2.0$. 



We now discuss the results of our experiments.  To give an indication of the evolution of a typical pathway over the course of a simulation, we show snapshots of the pathway DAG at various times of the first ensemble member from the Ex2 experiment for the 10 Tg eruption run (Figure \ref{fig:dagexample}).  Included in the supplementary materials are movies of the evolution of the pathway DAG of the first ensemble member from for 5,10, and 20 Tg eruption runs using Ex2.  Because the pathway DAG depends on the ensemble member, we characterize pathways in terms of the statistics of the pathway DAG across the ensemble.  In particular, we look at the first time QOIs are active and the total time QOIs are active in each pathway.  If QOI never becomes active, then its first activation time is set to day 1200 by convention.

\begin{figure*}
    \includegraphics[width=0.23\linewidth]{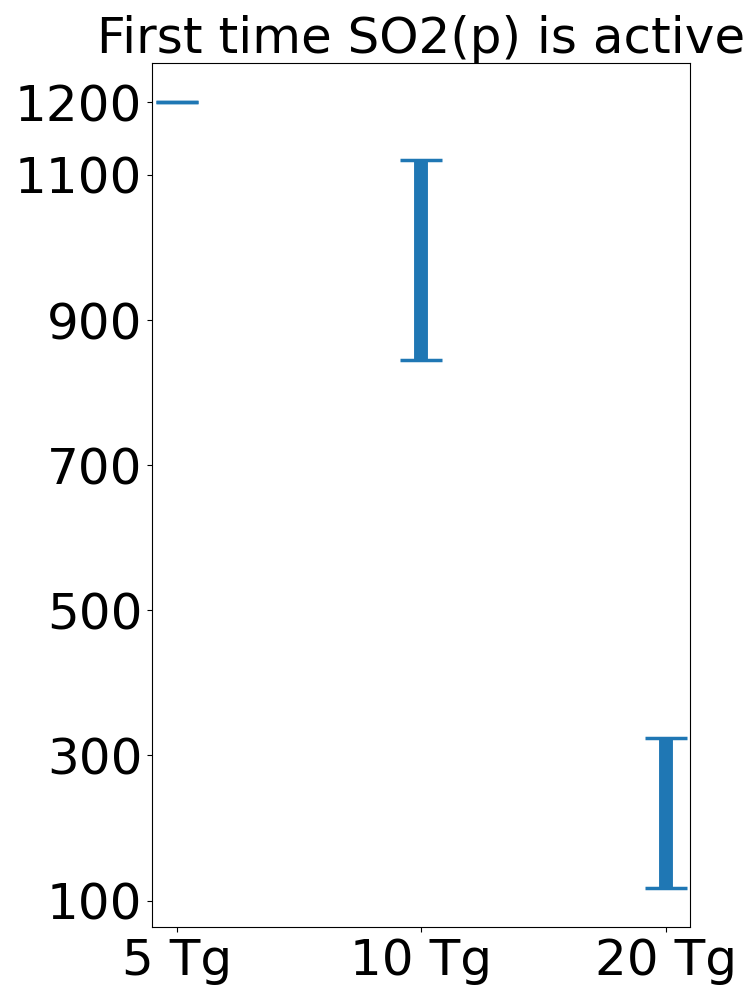}      \includegraphics[width=0.23\linewidth]{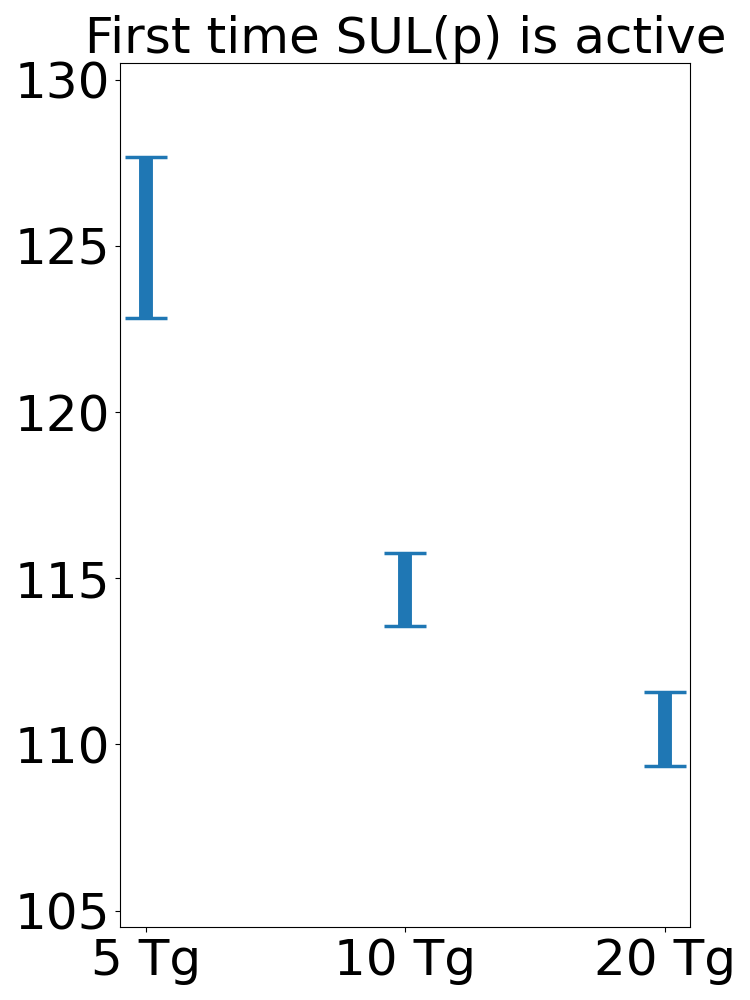}      \includegraphics[width=0.23\linewidth]{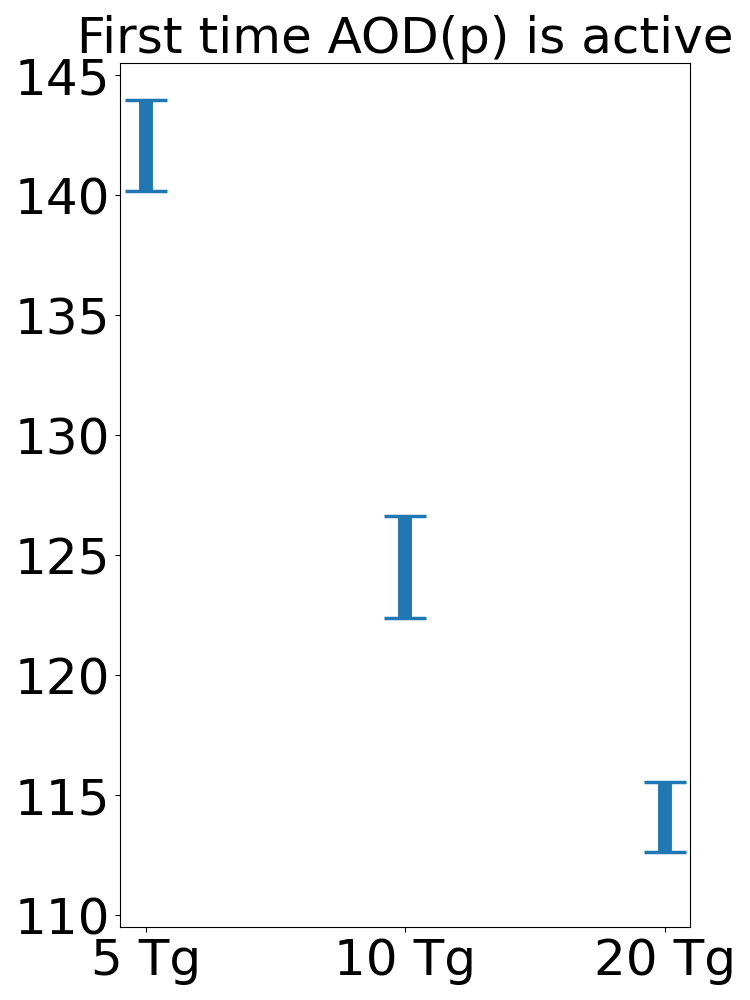}    \includegraphics[width=0.23\linewidth]{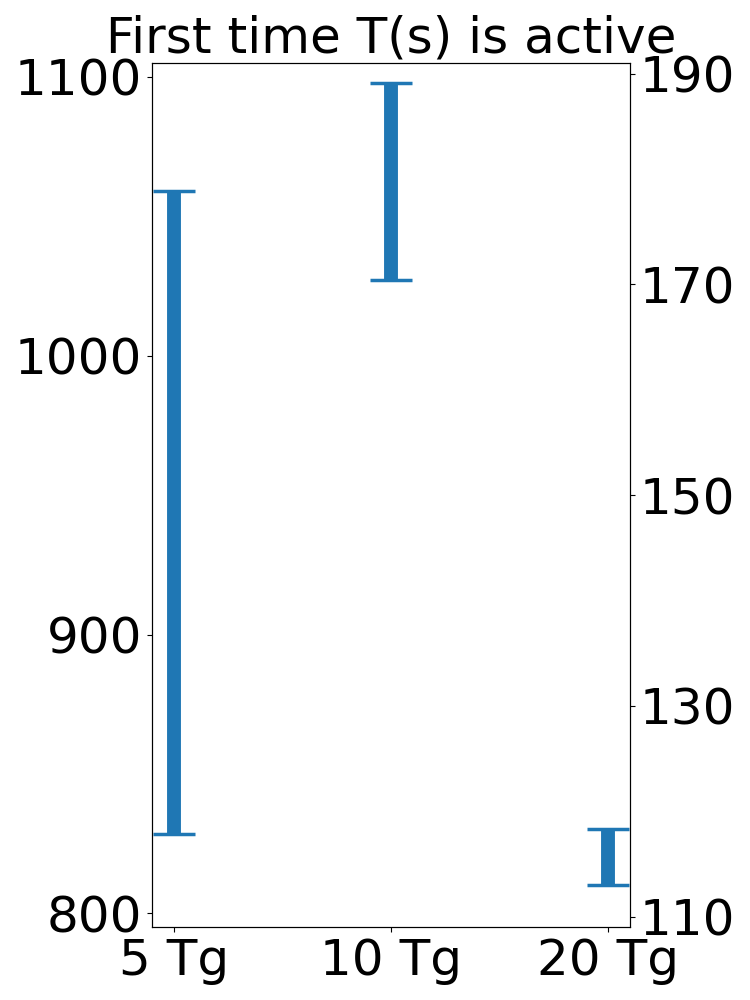}  \\
        \includegraphics[width=0.23\linewidth]{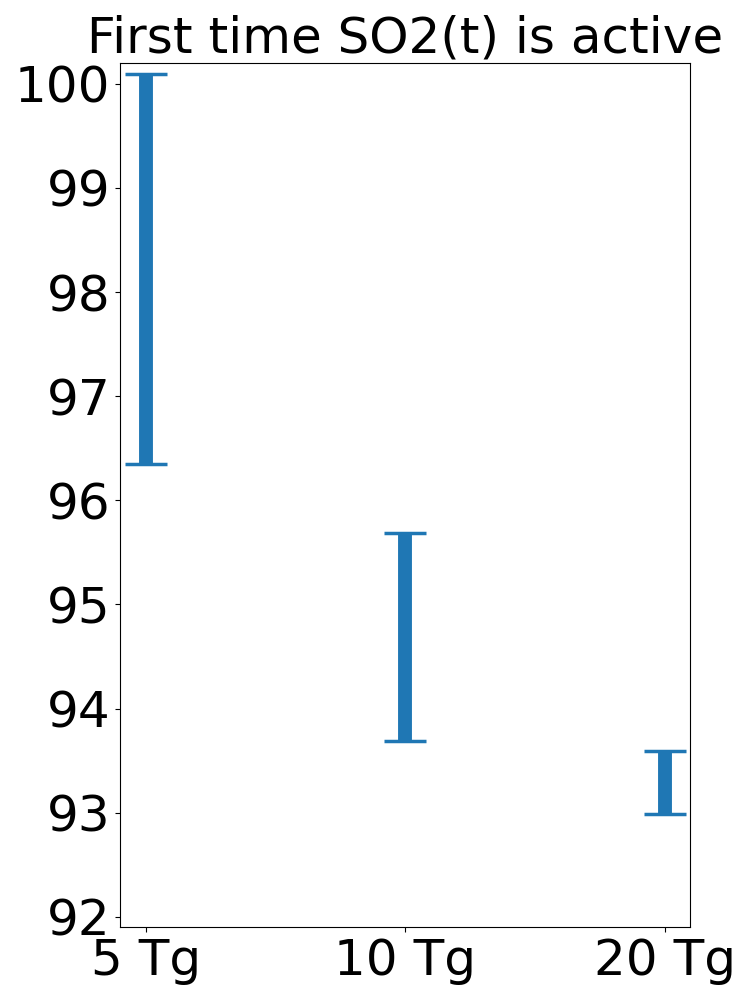}      \includegraphics[width=0.23\linewidth]{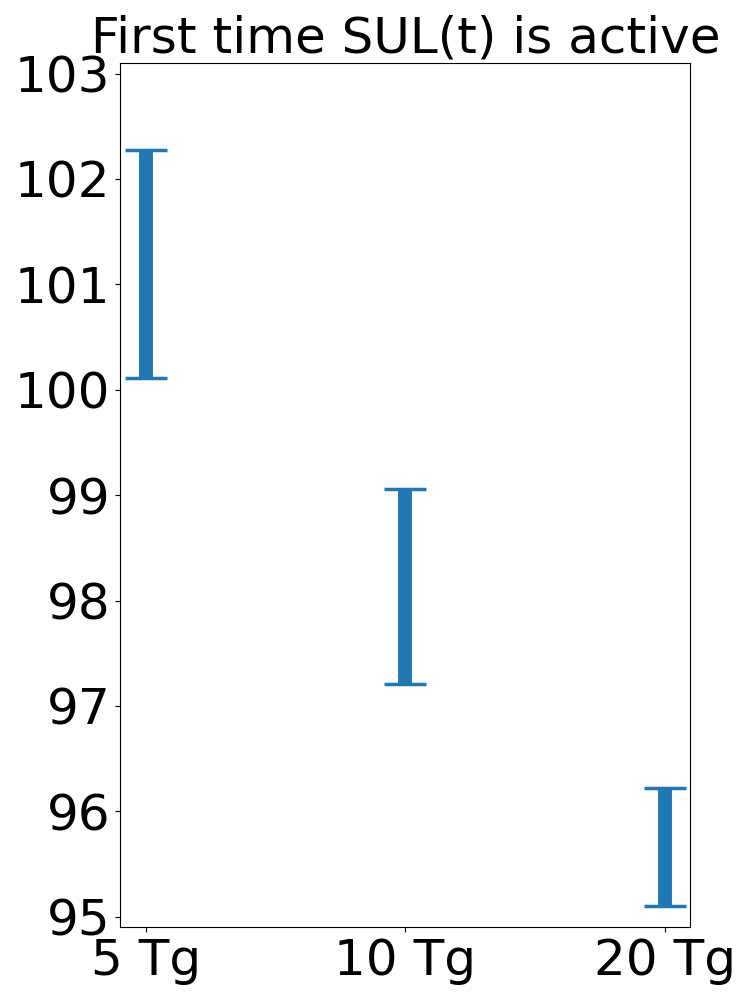}      \includegraphics[width=0.23\linewidth]{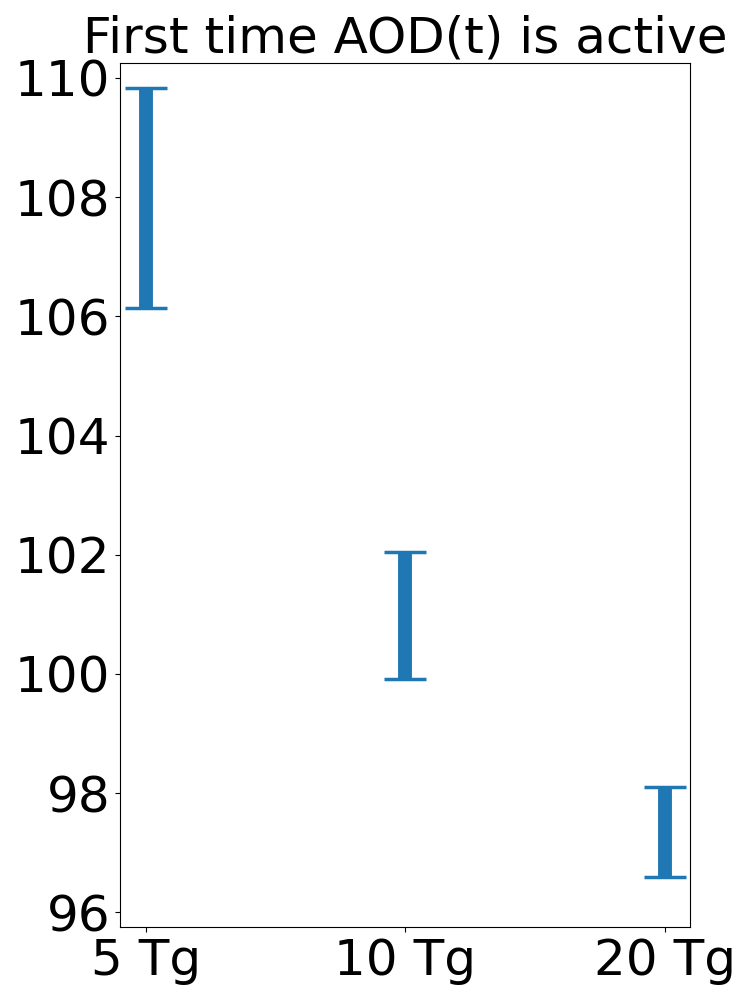}   
        \includegraphics[width=0.23\linewidth]{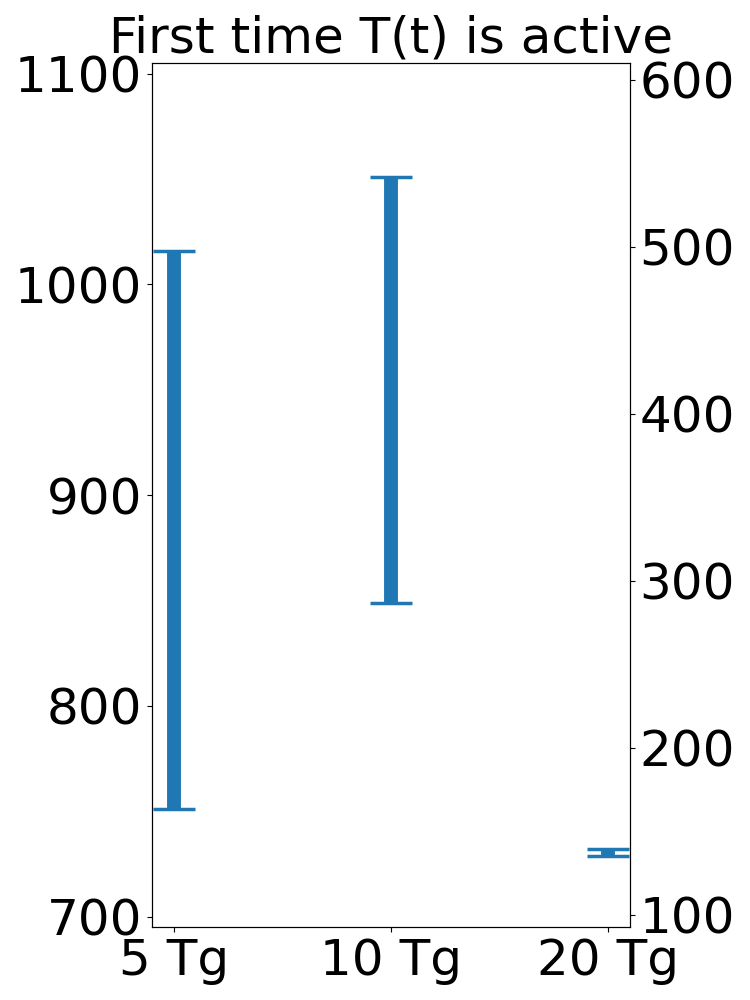}\\    \includegraphics[width=0.23\linewidth]{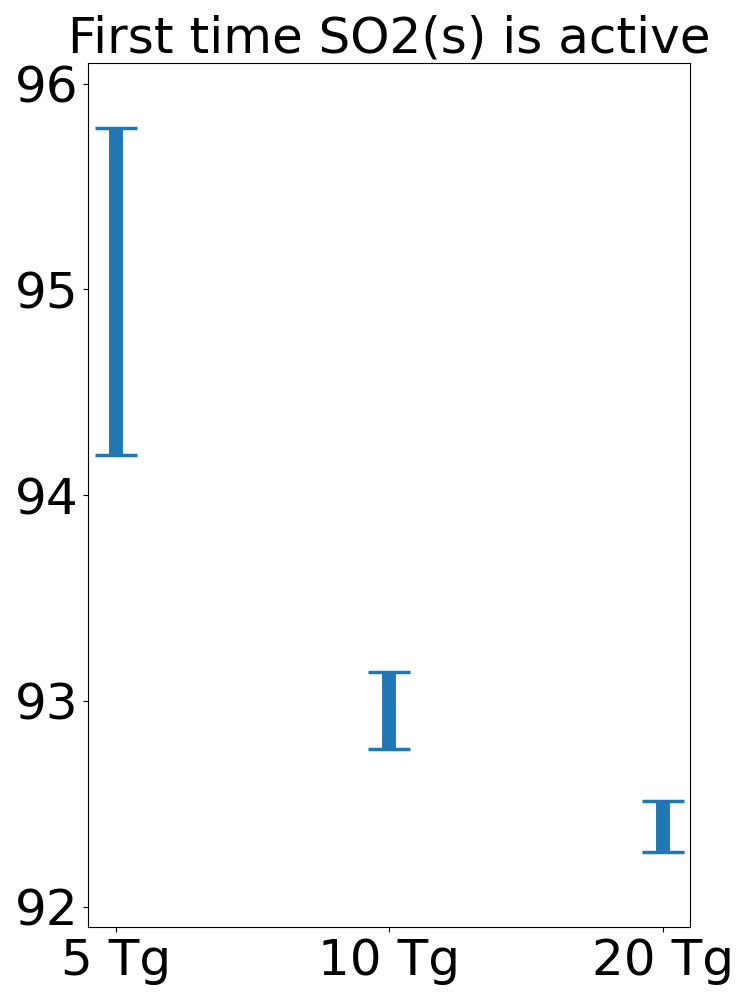}      \includegraphics[width=0.23\linewidth]{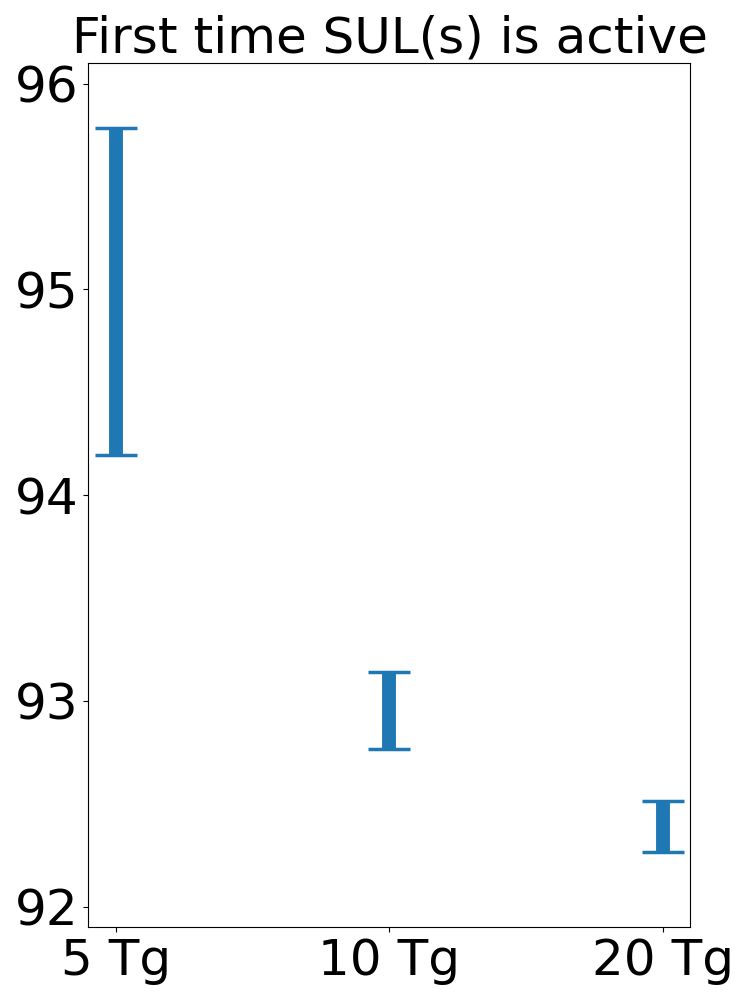}      \includegraphics[width=0.23\linewidth]{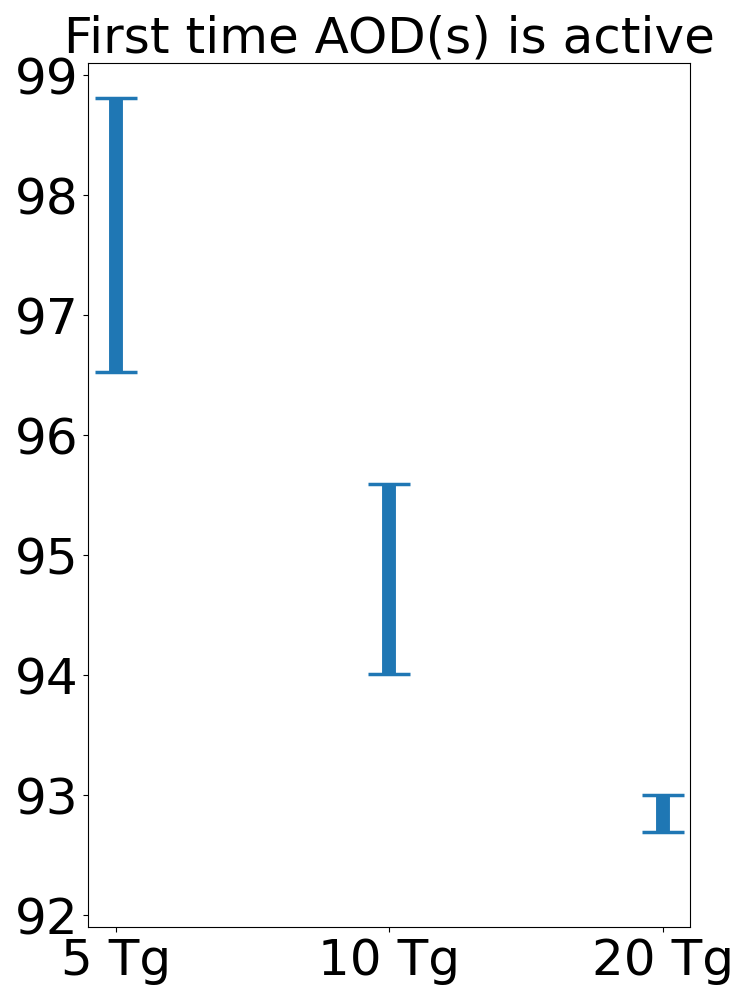}\includegraphics[width=0.23\linewidth]{figures/ft_T_s.png}\\
            \includegraphics[width=0.23\linewidth]{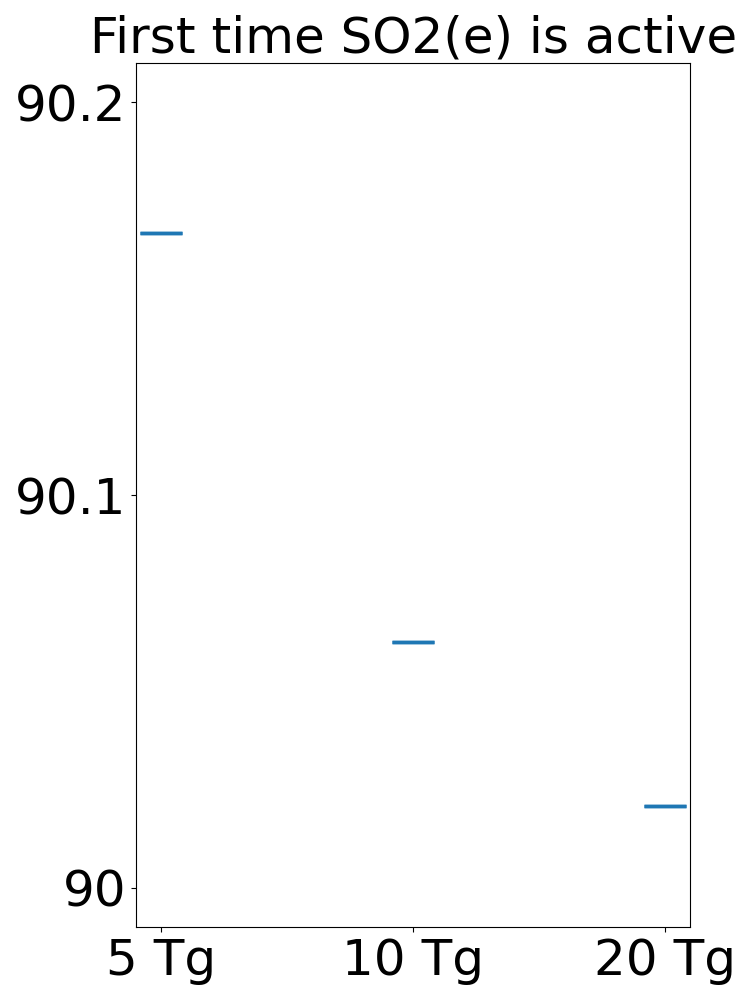}      \includegraphics[width=0.23\linewidth]{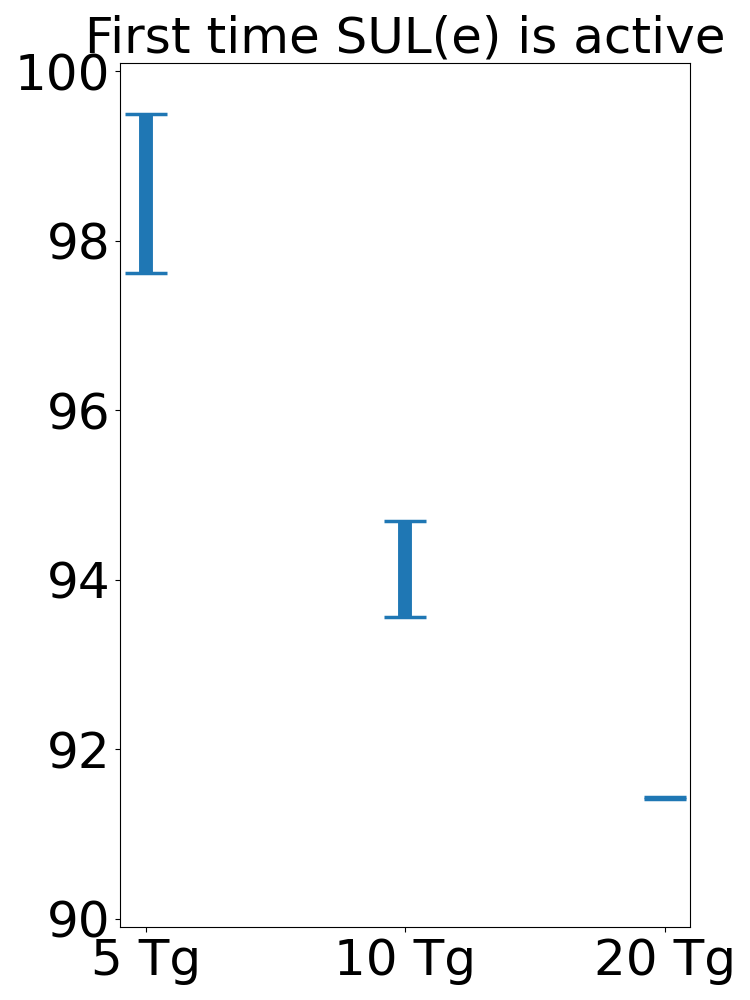}      \includegraphics[width=0.23\linewidth]{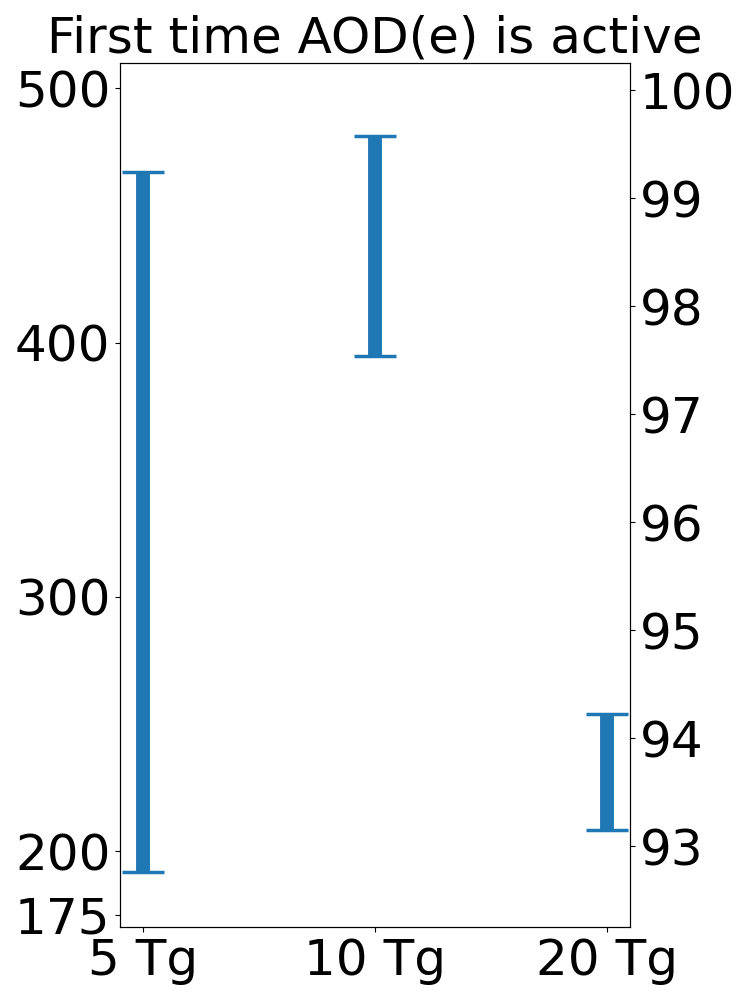}
            \includegraphics[width=0.23\linewidth]{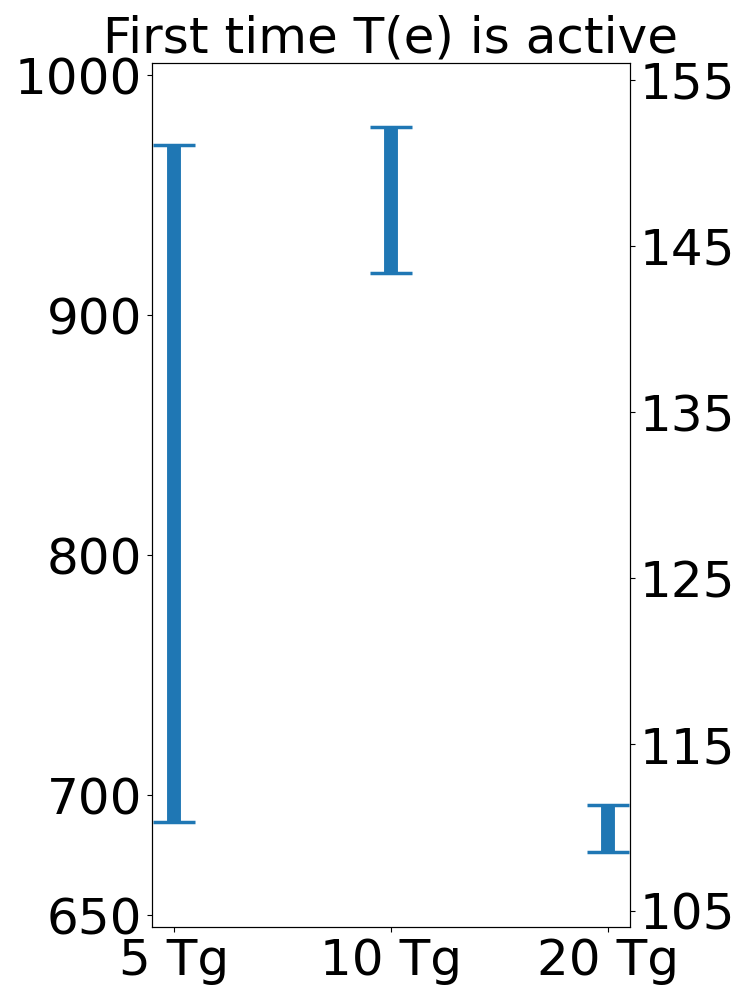}
    \caption{First activation times with error bars denoting the standard error (mean and standard error taken with respect to the 10-element ensemble corresponding to an eruption mass of 5,10,20 Tg) for SO2, SUL, AOD, T (using the Ex2 bounds test) in all four zones.  In figures with two y-axes, the left y-axis on the left corresponds to the 5 Tg eruption mass and the right y-axis on the right corresponds to the 10 and 20 Tg eruption masses.}

    \label{fig:firsttime}
\end{figure*}

\begin{figure*}
    \includegraphics[width=0.23\linewidth]{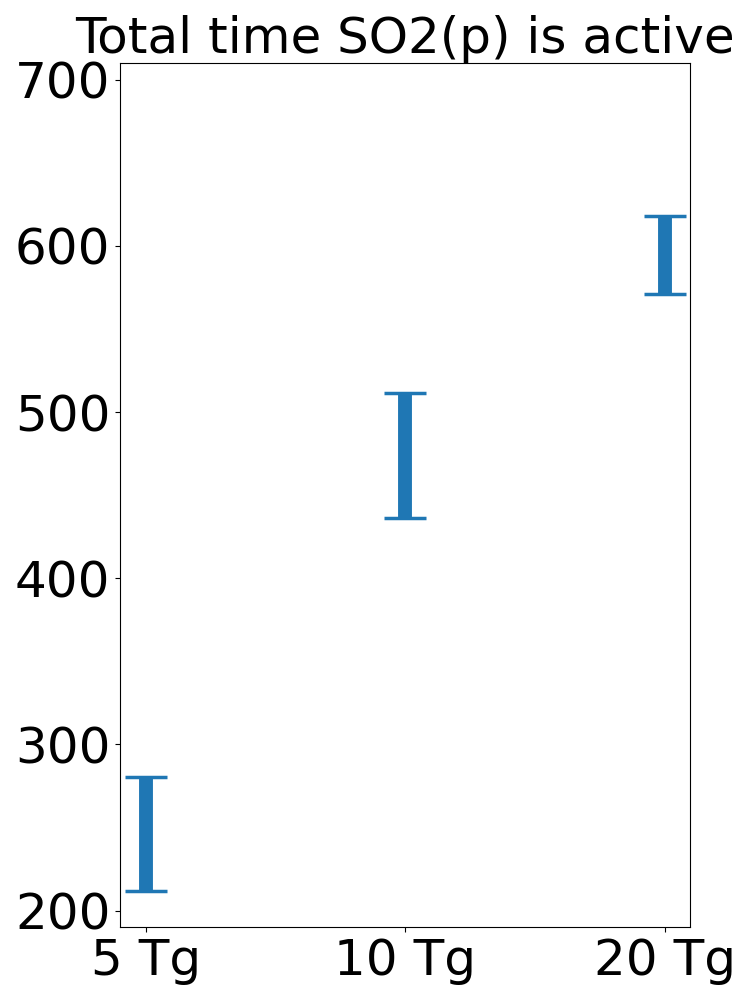}      \includegraphics[width=0.23\linewidth]{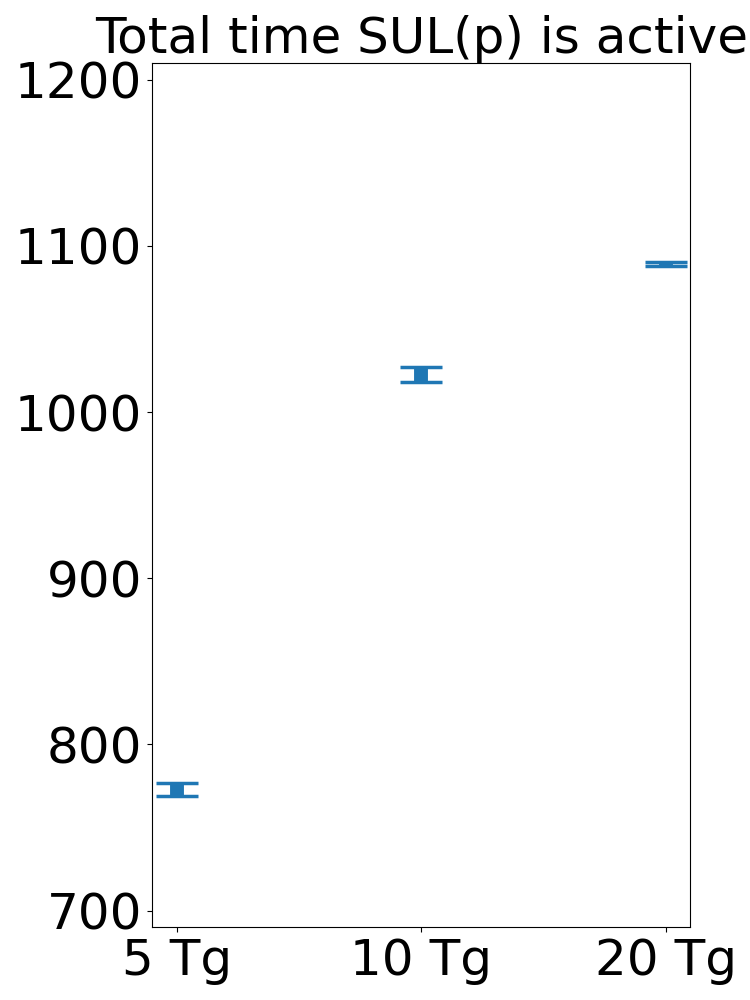}      \includegraphics[width=0.23\linewidth]{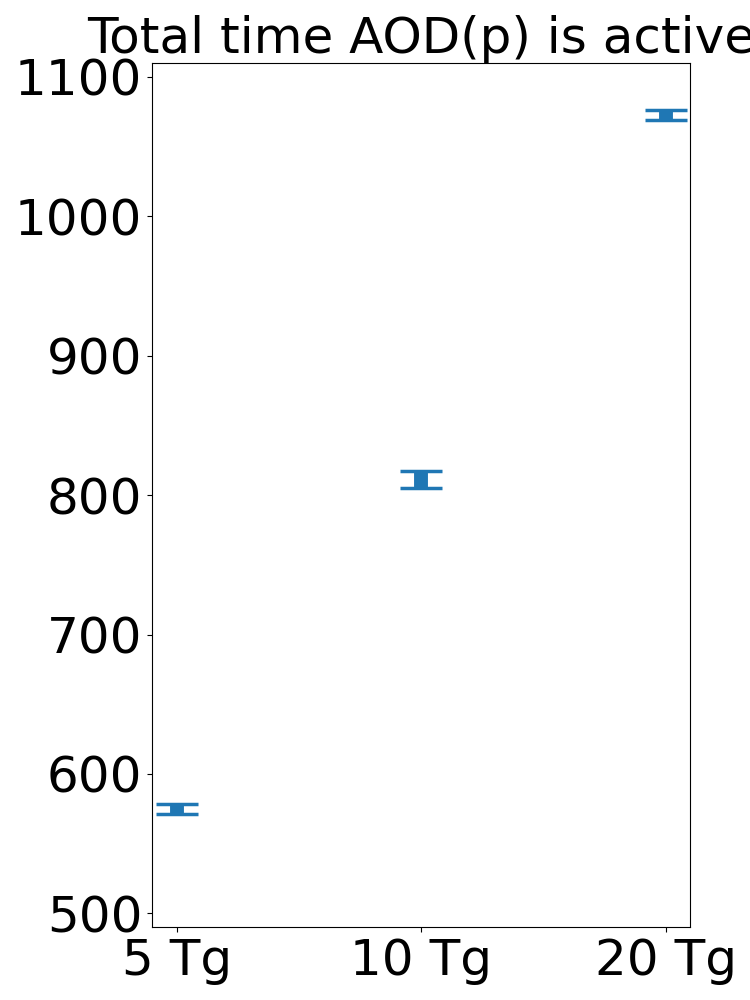}    \includegraphics[width=0.23\linewidth]{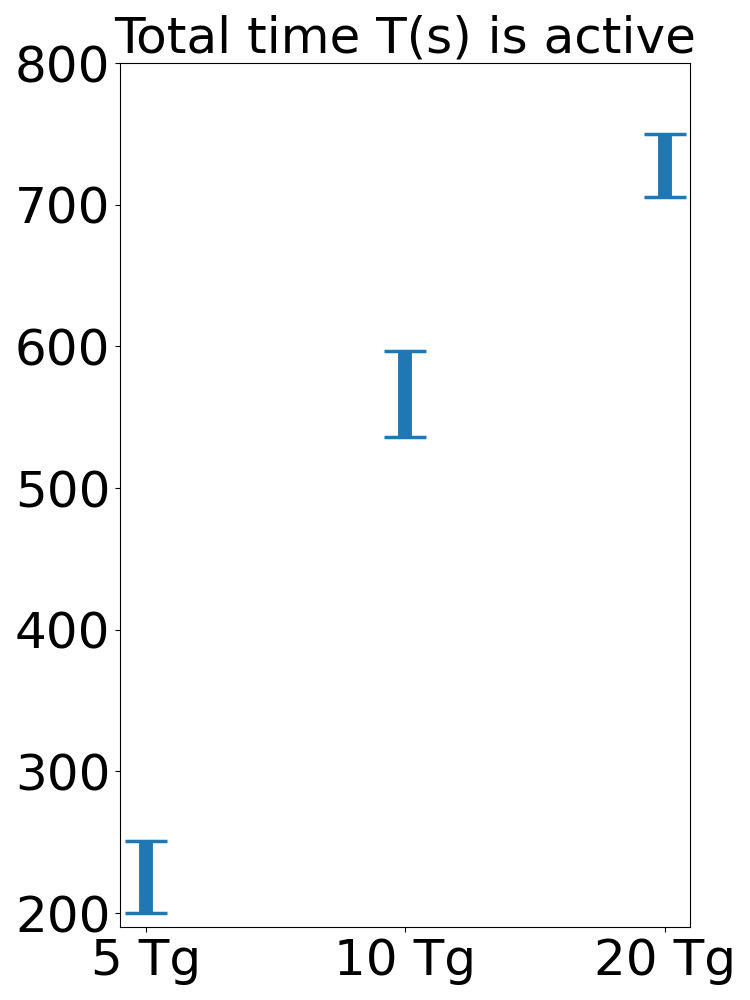}  \\
        \includegraphics[width=0.23\linewidth]{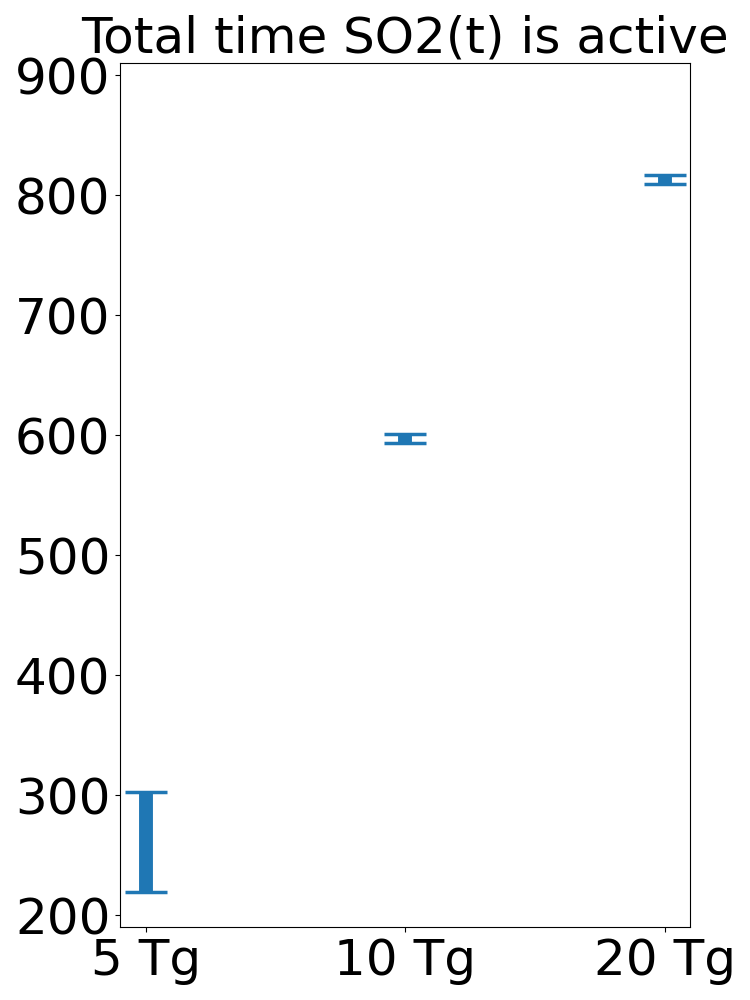}      \includegraphics[width=0.23\linewidth]{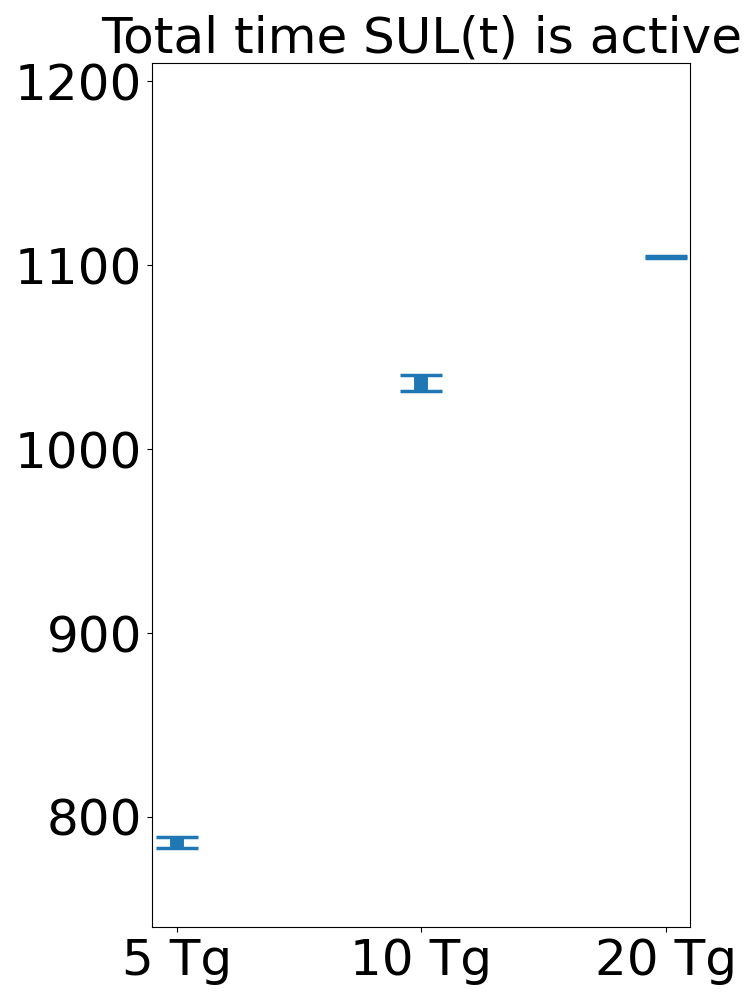}      \includegraphics[width=0.23\linewidth]{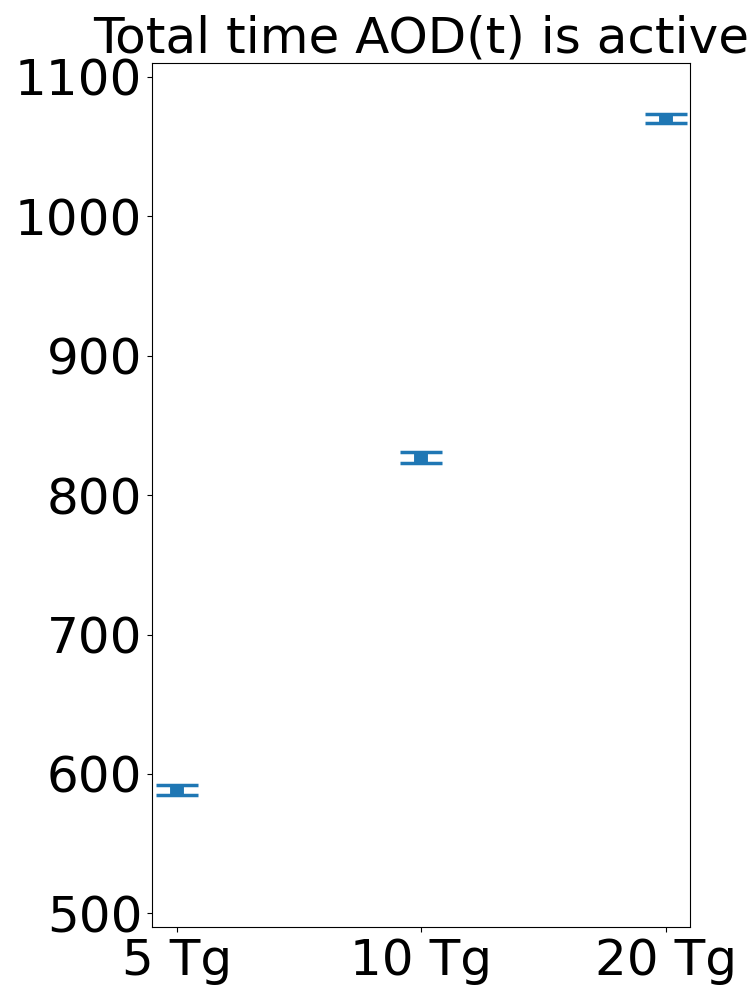}   
        \includegraphics[width=0.23\linewidth]{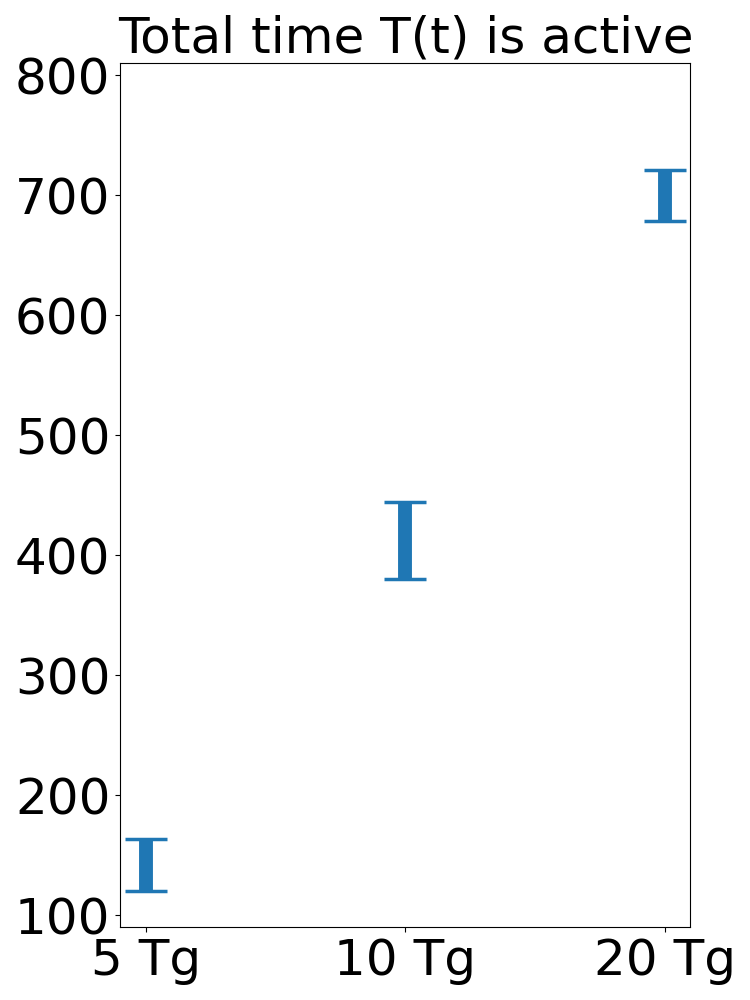}\\    \includegraphics[width=0.23\linewidth]{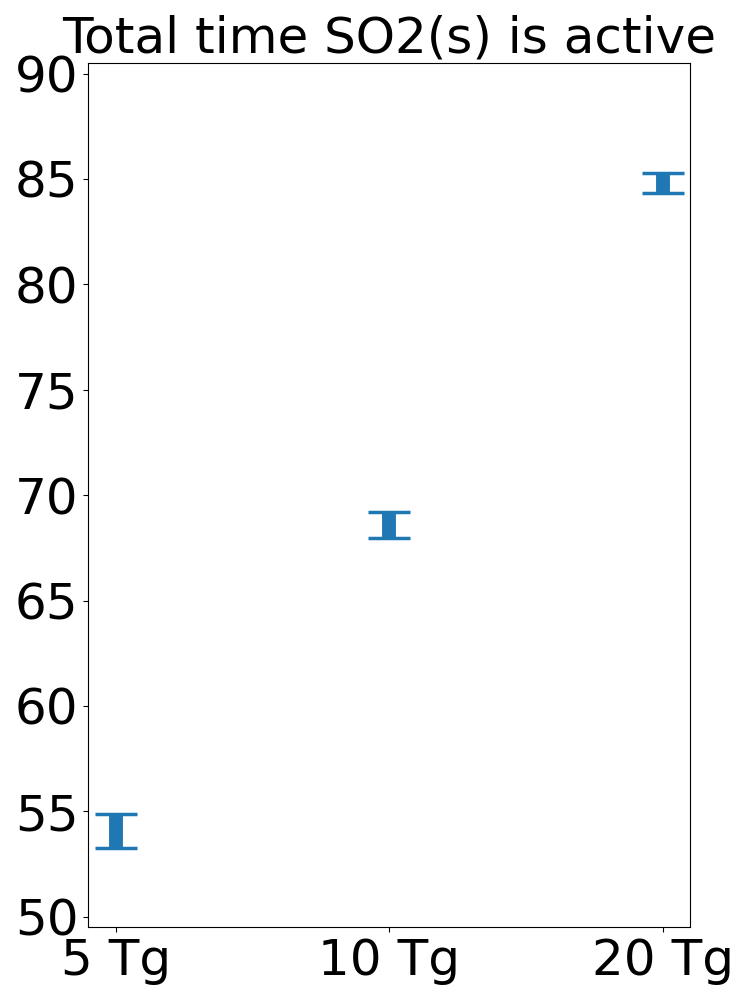}      \includegraphics[width=0.23\linewidth]{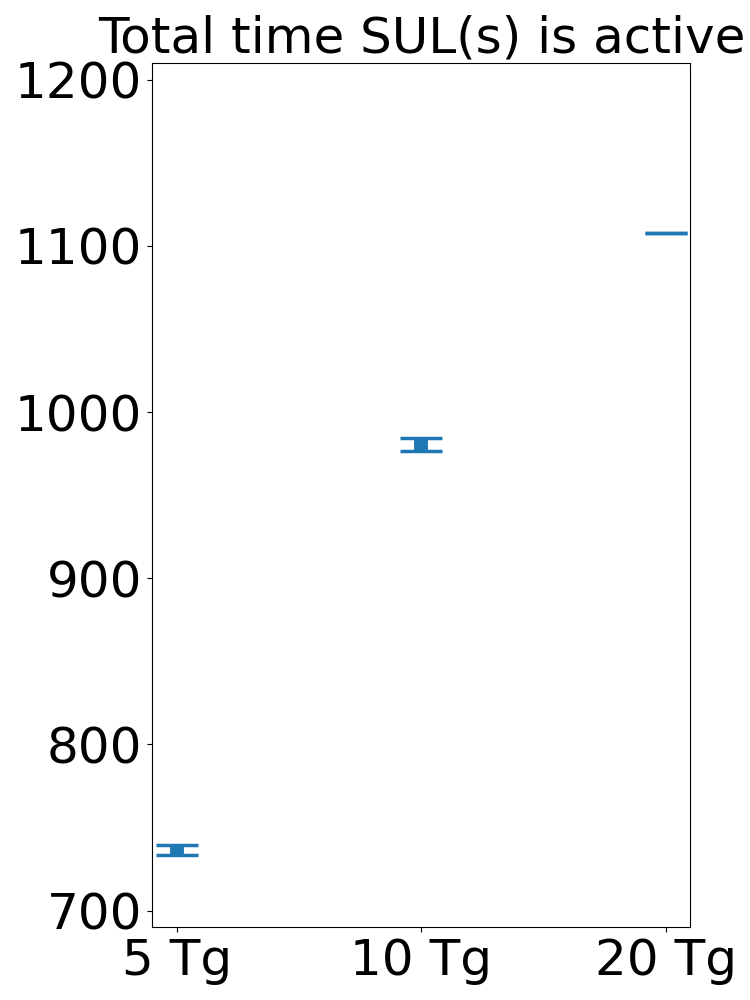}      \includegraphics[width=0.23\linewidth]{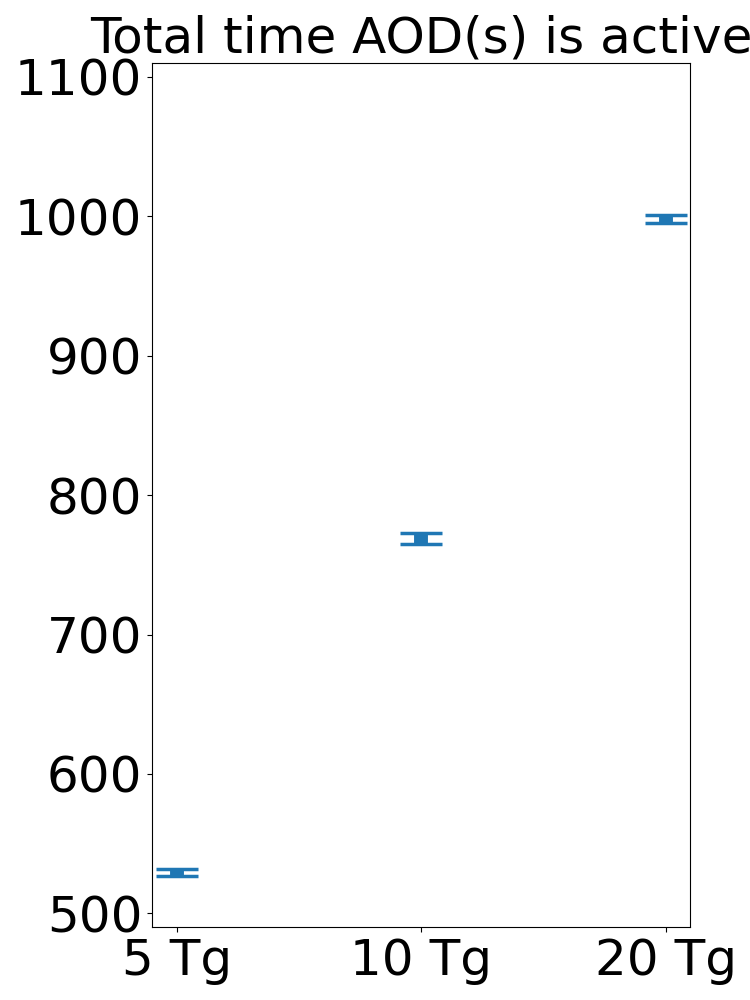}\includegraphics[width=0.23\linewidth]{figures/tt_T_s.png}\\
            \includegraphics[width=0.23\linewidth]{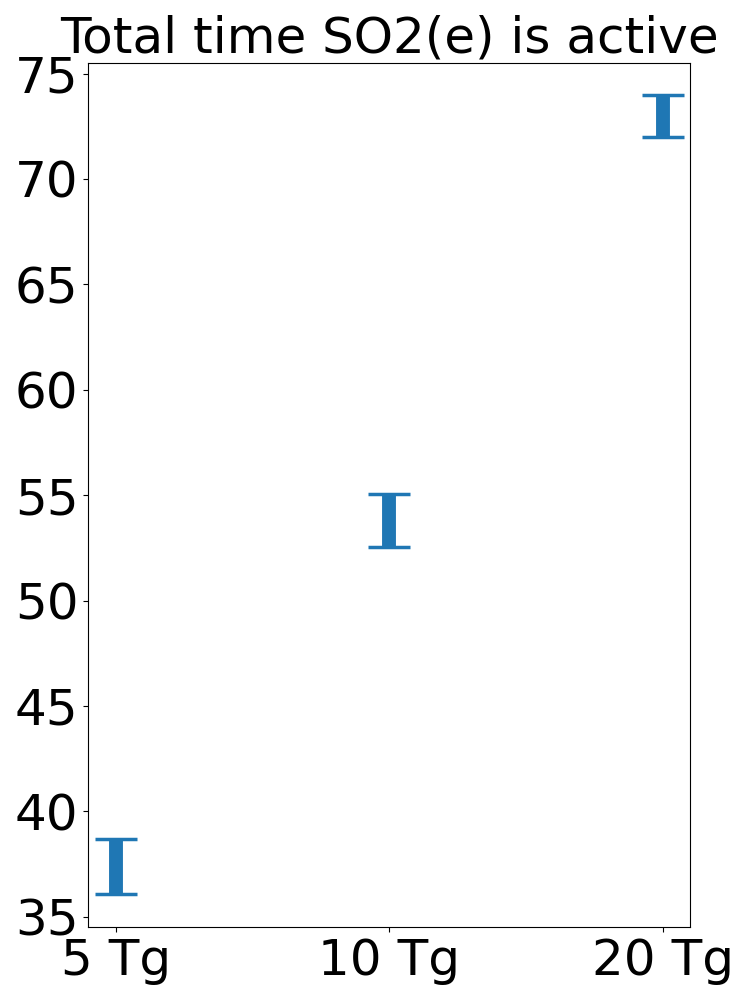}      \includegraphics[width=0.23\linewidth]{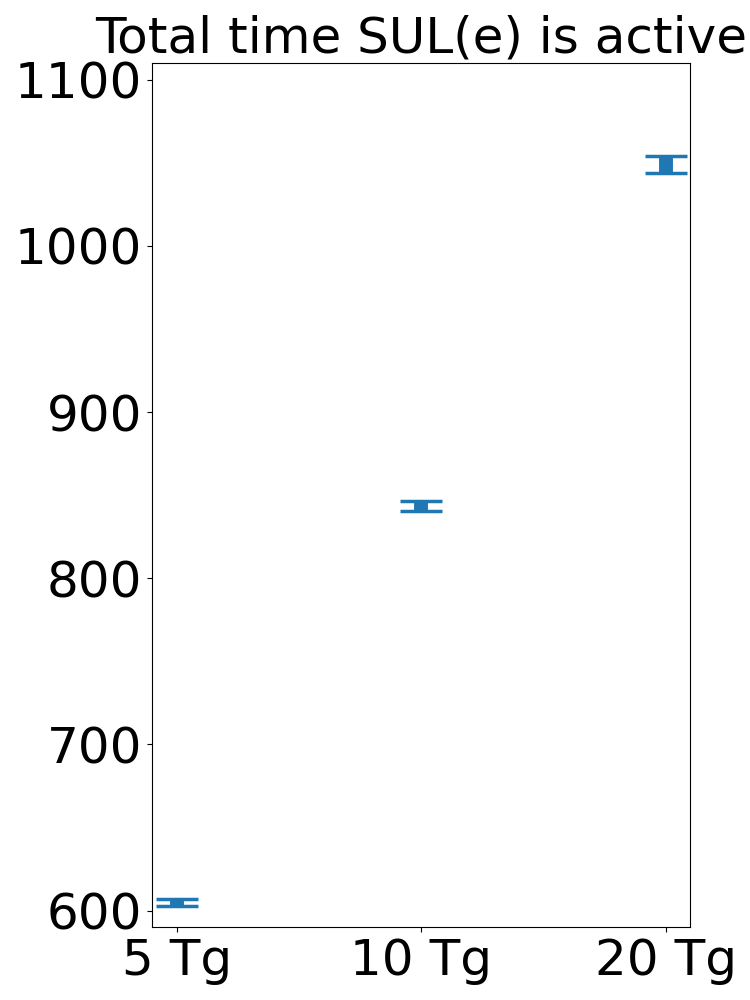}      \includegraphics[width=0.23\linewidth]{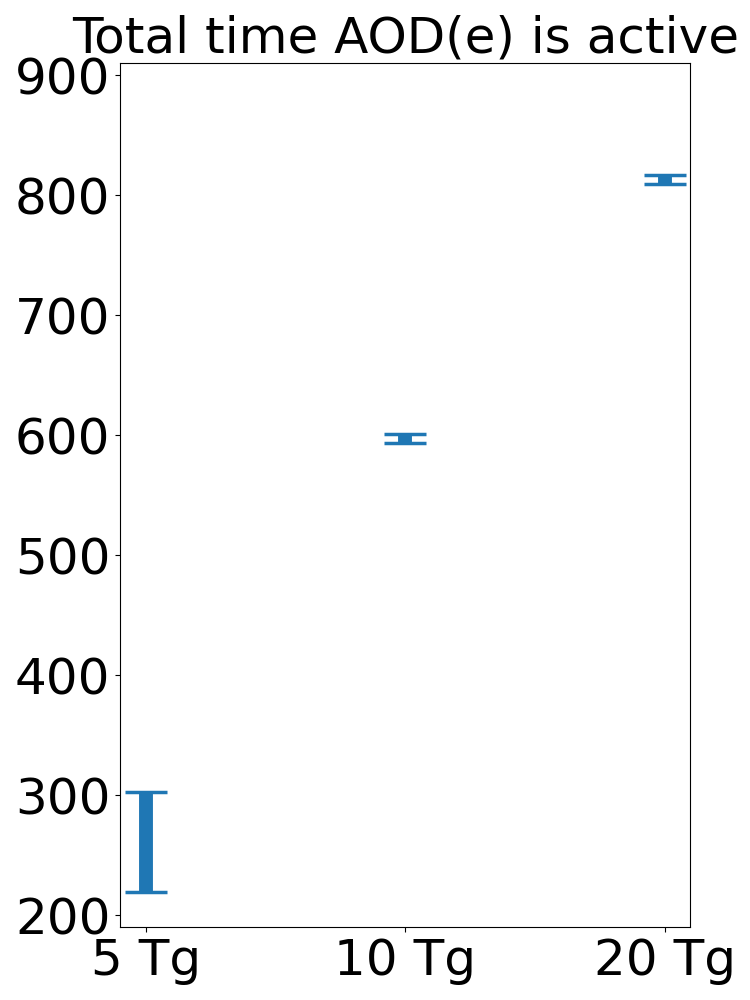}
            \includegraphics[width=0.23\linewidth]{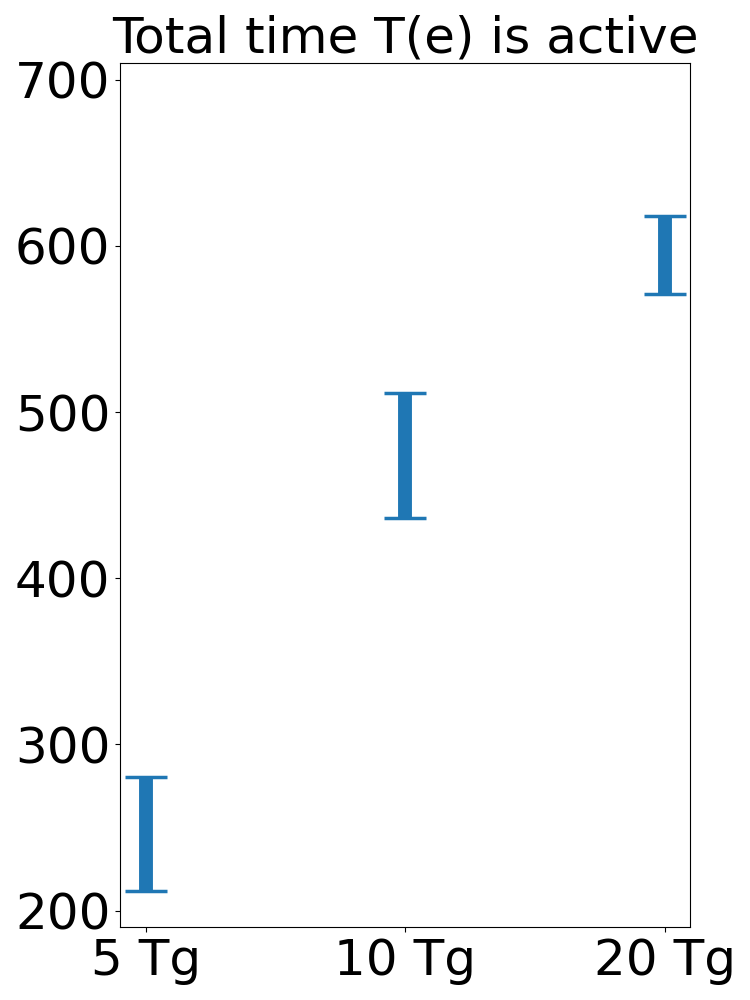}
    \caption{Total activation times with error bars denoting the standard error (mean and standard error taken with respect to the 10-element ensemble corresponding to an eruption mass of 5,10,20 Tg) for SO2, SUL, AOD, T (using the Ex2 bounds test) in all four zones.  }

    \label{fig:totaltime}
\end{figure*}

\begin{figure*}
    \includegraphics[width=0.5\linewidth]{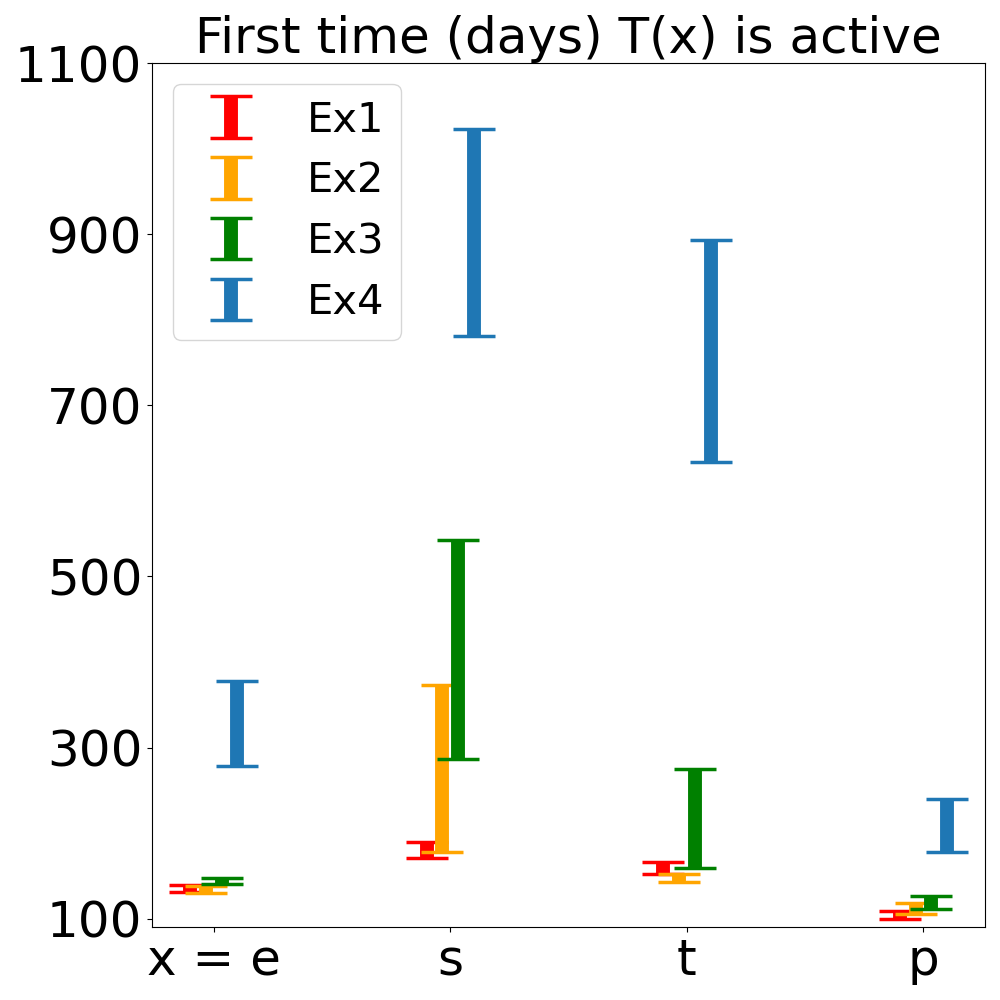}\includegraphics[width=0.5\linewidth]{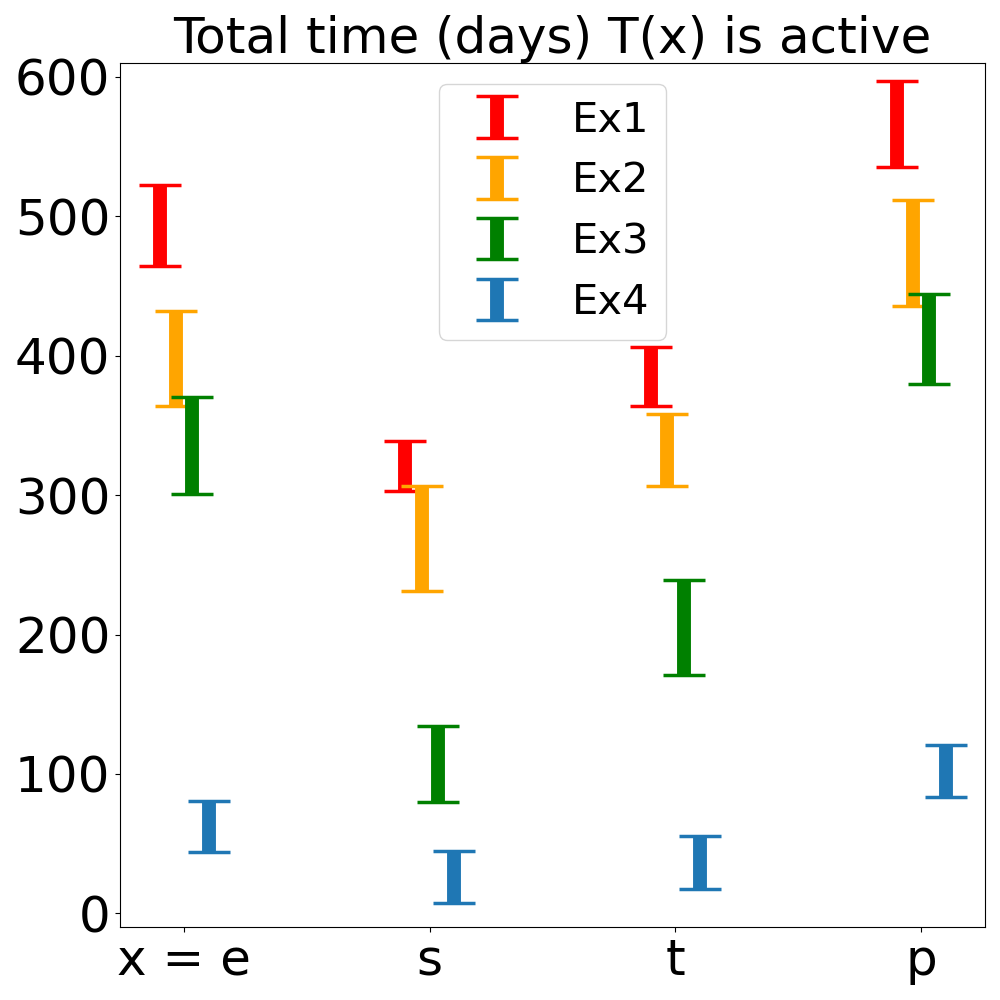}
\caption{(Left) First time T(x) is active in the 10 Tg eruption ensemble in Ex1,Ex2,Ex3,Ex4. (Right) Total time T(x) is active in the 10 Tg eruption ensemble in Ex1,Ex2,Ex3,Ex4.}
    \label{fig:Texexperiment}
\end{figure*}

In Figure \ref{fig:firsttime}, we plot the mean first time at which each QOI is active in each pathway-DAG. Figure \ref{fig:totaltime} plots the total time over which QOI is active.  Figure \ref{fig:Texexperiment} shows the effect of varying $T_u$ in Ex1-Ex4 on the first active time and total active time of $T(x)$ for the 10 Tg eruption.  The mean values are computed across each 10-element ensemble and the bars at the end of the vertical lines denote the corresponding standard error.  Across all QOIs, the broad trend is that variables are active sooner and for a longer span of time in runs with a larger eruption mass.  However, there is still quite a bit of variability, particularly (see Figure \ref{fig:firsttime}) for the first activation time in the 5 Tg eruption ensemble for AOD(e) and T.  We also note that experiments which are more sensitive (smaller values of $T_u$ correspond to T(x) activating sooner and for longer times as would be expected (see the bottom right plots of Figures \ref{fig:Texexperiment}).  We note that SO2(p) rarely becomes active during a simulation for any of the eruption runs since most SO2 has converted into SO4 by the time it was been transported to the poles.

Figure \ref{fig:dagexample} shows snapshots of a DAG from a 10 Tg eruption run at different times, highlighting it as a subset of the base DAG. Examining the snapshots from left to right and top to bottom shows an ``activation wave'' propagating from the bottom left to the top right. The wave highlights how the anomaly in the QOIs develops first in the equatorial region (where the eruption happens) and then propagates towards the polar region via the Brewer-Dobson circulation, as expected. In addition to spatial propagation, the evolution of the DAG also highlights the propagation of the signal across QOIs within the same region, in accordance with the analytically programmed relationships in HSW-V for how SO2 impacts temperature via changes in SUL and AOD.  We emphasize again that the pathway DAG does not in itself establish causality; however, it corroborates the physics- and observation-driven intuition that anomalously large values in SO2 concentrations result in anomalies in SUL, AOD, and eventually T.

\newpage
\newpage


\begin{figure*}
  \centering

   \includegraphics[width=0.4\linewidth]{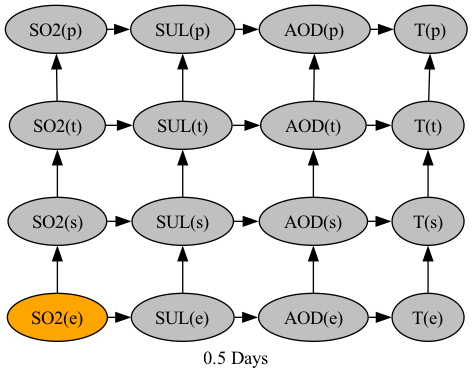}\vline\includegraphics[width=0.4\linewidth]{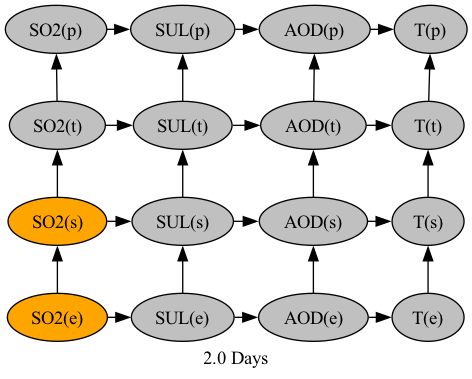}
        \vspace{0.05cm}

    \includegraphics[width=0.4\linewidth]{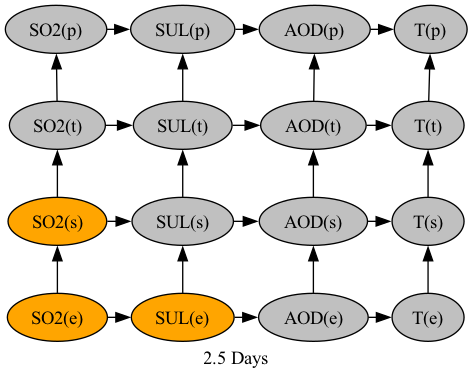}\vline \includegraphics[width=0.4\linewidth]{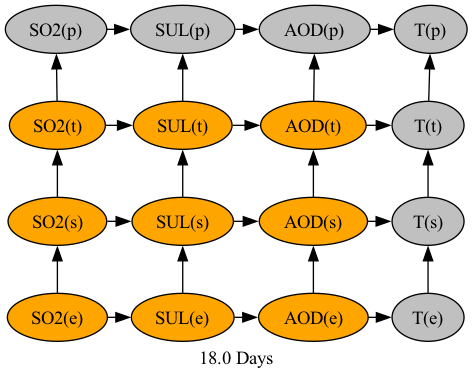}
      \vspace{0.05cm}
      
   \includegraphics[width=0.4\linewidth]{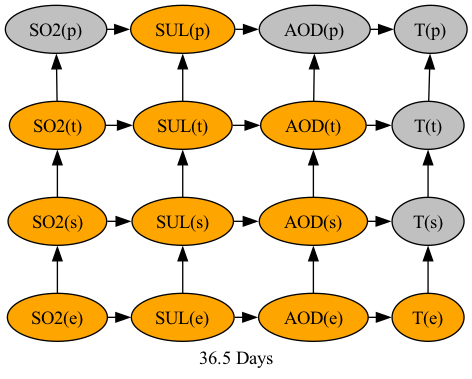}\vline\includegraphics[width=0.4\linewidth]{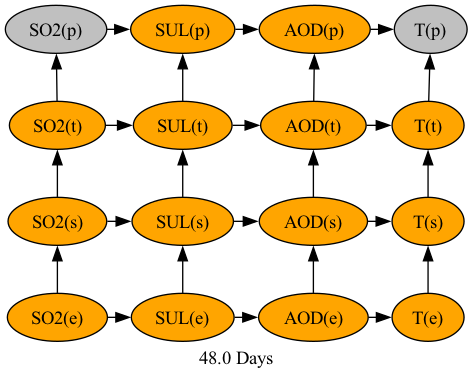}
   \vspace{0.05cm}

   \includegraphics[width=0.4\linewidth]{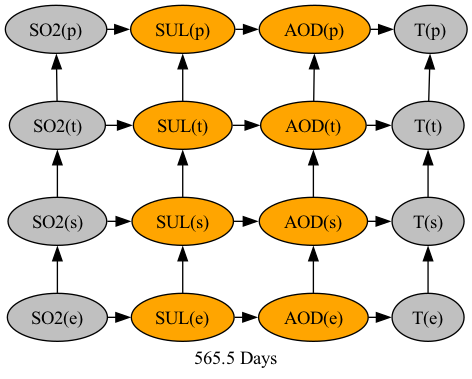}\vline\includegraphics[width=0.4\linewidth]{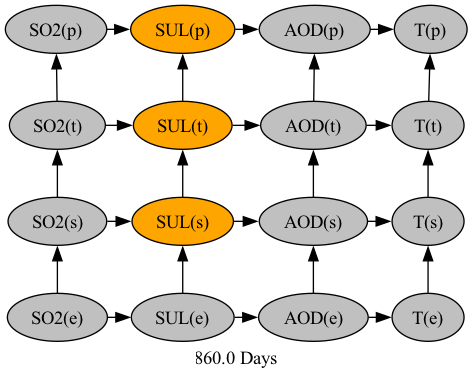}
 
    \caption{Snapshots of the pathway DAG of the first element of the Ex1 10Tg ensemble at various post-eruption times where active nodes are shaded orange and inactive nodes are shaded gray.}
  \label{fig:dagexample}
\end{figure*}


%% file: sections/6_conclusion.tex
We have presented a method for characterizing and analyzing source-impact pathways and the \ct{} software package for extracting the necessary data in-situ.  In particular, we have shown that this analysis can be done efficiently and that our algorithm correctly captures impacts of the source-impact temperature pathway of a volcanic eruption simulated in HSW-V.  A next step will be source-impact pathway analysis in more complex and realistic \esm{} configurations. Studying the impacts of volcanic eruptions on more than just stratospheric temperature will require extending \ct{} to model components beyond the atmosphere.  We are currently extending \ct{} to the E3SM Land Model (ELM), towards understanding the impact of volcanic eruptions on agriculture and crop production.


Another application of \ct{} would be in-situ monitoring and real-time manipulation of internal model variables on the native model grid.  Modification of internal model variables would enable studies of the effects of geoengineering or climate intervention strategies.  Additionally, this could be used for causality analysis that could subsequently be used to create pathway DAGs where the direct edges indicate some notion of causality.






